# Bloch Oscillations in Chains of Artificial Atoms Dressed with Photons

Ilay Levie and Gregory Slepyan[*]

*School of Electrical Engineering, Tel Aviv University, Tel Aviv 39040, Israel*

We present the model of one-dimensional chain of two-level artificial atoms driven with dc field and quantum light simultaneously in strong coupling regime. The interaction of atoms with light leads to entanglement of electron and photon states (dressing of the atoms). The driving via dc field leads to the Bloch oscillations (BO) in the chain of dressed atoms. We considered the mutual influence of dressing and BO and show that scenario of oscillations dramatically differs from predicted by the Jaynes-Cummings and Bloch-Zener models. We study the evolution of the population inversion, tunneling current, photon probability distribution, mean number of photons, photon number variance and show the influence of BO on the quantum-statistical characteristics of light. For example, collapse-revivals picture and vacuum Rabi-oscillations are strongly modulated with Bloch frequency. As a result, quantum properties of light and degree of electron-photon entanglement become controllable via adiabatic dc field turning. On the other hand, the low-frequency tunneling current depends on the quantum light statistics (in particular, for coherent initial state it is modulated accordingly the collapse-revivals picture). The developed model is universal with respect to the physical origin of artificial atom and frequency range of atom-light interaction. The model is adapted to the 2D-heterostructures (THz frequencies), semiconductor quantum dots (optical range), and Josephson junctions (microwaves). The data for numerical simulations are taken from recently published experiments. The obtained results open a new ways in quantum state engineering and nano-photonic spectroscopy.

## I. INTRODUCTION

The early quantum theory of electrical conductivity in crystal lattices by Bloch, Zener and Wannier [1-4] led to the prediction that a homogeneous dc field induces an oscillatory rather than uniform motion of the electrons. These so-called Bloch oscillations (BO) have been observed in bulk n-doped GaAs at a lattice temperature 300K in high fields up to 300 kV/cm [5] as well as different types of artificial systems such as semiconductor superlattices [6], interacting atoms in optical lattices [7,8], ultracold atoms [9–14], light intensity oscillations in waveguide arrays [15-21], acoustic waves in layered and elastic structures [22], atomic oscillations in Bose-Einstein condensates [23], among others. Several recent studies have investigated the dynamics of cold atoms in optical lattices subject to ac forcing; the theoretically predicted renormalization of the tunneling amplitudes has been verified experimentally. The recent observations include global motion of the atom cloud, such as giant "super–Bloch oscillations" [24]. As a result, BO transformed from the specific contraintuitive model to the general experimentally supported physical concept of oscillatory motion of wave packets placed in a periodic potential when driven by a constant force [8,25].

Rabi oscillations (RO) are periodical transitions of a two-level quantum system between its stationary states under the action of ac driving field [26, 27]. The phenomenon was theoretically predicted by Rabi for the nuclear spins in radio-frequency magnetic field [28] and afterwards, discovered in various physical systems, such as electromagnetically driven individual atoms [29] including the case of Rydberg atomic states [30], semiconductor quantum dots (QDs) [31] and different types of solid-state qubits (superconducting charge qubits based on Josephson junctions [32-34], spin qubits [35], semiconductor charge qubits [36]). Text-book picture of Rabi effect is given by the Jaynes-Cummings model [26, 27]. It implied the concept of the dressed atom with light, correspondent to the quantum entanglement of electrons and photons. This model can be essentially modified by a set of additional features, such as the broken inversion symmetry [37], the propagation of RO over the chains of coupled atoms in the form of special waves (Rabi-waves) and depolarization due to the local fields [38-43]. It opens the possibility of generation a variety of entangled quantum states, which could have a great impact on the search for universal and efficient quantum computation processes [44], the electrically tunable optical nano-antennas with highly directive emission [39,40,43], quantum sensing [45], and metrology [46]. Recent theoretical progress in this area is associated, in particular, with the novelty of inter-atomic coupling mechanisms [48], which manifests themselves in surprising thermodynamic behavior of the specific waveguiding arrays, which recently was experimentally verified [49].

In this paper, we build a theoretical model of a chain of coupled two-level quantum elements exposed to the quantum light and driven via bias voltage. We consider the case of strong coupling of light with charge carriers, which leads to the entanglement of electron-photon quantum states. This model describes BO of electrons dressed with light and their mutual influence with RO. Our model has a significant degree of generality: it relates to the systems of different physical origin and various frequency ranges. We consider its application to semiconductor heterostructures (THz frequencies), semiconductor quantum dots (visible frequencies), Josephson junctions (microwaves). For brevity, we refer to every of these artificial two-level quantum objects as an "atom" regardless of its physical implementation.

We develop the model taking into account conditions of real experiments. For example, the coherent intersubband excitation of heterostructure in THz region has been done in [50] by the ultrashort (femtosecond) pulses. On this account, we generalized our model for the case of electromagnetic pulse, advancing the secondary quantization of fields to the case of pulses. Our model is based on some conventional simplifications. We use rotating-wave approximation (RWA) and neglect any damping. This requires the fulfillment of certain relations between the frequencies (transmission frequency, light frequency, Rabi-frequency, Bloch frequency) and between some characteristic times (coherence time, attenuation time). Our calculations have been made for real physical parameters of atoms and their environment, and accessibility of these relations have been supported by the achievements of modern technologies and the data of published experiments. It allows the design of installations for potential future experiments in this branch.

The classical analog of our investigation is Rabi-Bloch oscillations (RBO) predicted in [51,52]. The quantum origin of light makes the subject dramatically changing. The classical light in the RBO case does not undergo a reverse reaction from electrons due to their transitions from lower level to upper one and vice versa. Therefore, the light plays role of the effective refractive medium for electron wavepacket, which guides the spatial propagation of Rabi-wave [38-41]. For the case of quantum light, the quantum electron transitions are accompanied by time-by-time emission-absorption of photons. Therefore, it is impossible to consider the individual electrons in the capacity of consistent Bloch oscillators. The main aim of the present paper is the analysis of mutual influence of RO and BO for the case of quantum light basing on the fundamental principles of quantum optics [26]. Exactly, electrons dressed with photons compose the type of quasi-particles, which are helpful for description of the considered complicated dynamics, using the picturesque BO language. The main novel predictions are: (i) the electron-photon entanglement is modulated with Bloch frequency via BO in dc field; (ii) the motion of quasiparticle as a comprehensive whole is modulated with Rabi frequency via interband optical transitions. We hope that these basical results will be able to stimulate the statement of new experiments.

The paper is organized as follows. In Sect. II we review the model and basic assumptions, formulate Hamiltonian at the Wannier-Fock basis and obtain equations of motion for probability amplitudes. In Sect. III we obtain the approximate analytical solution of equations of motion basing on quasi-classical concept. In Sect. IV we describe and discuss the results of numerical calculations for electron Gaussian wave packets and different initial states of light (coherent state, vacuum field, double-Fock state entangled with Gaussian wave packet). In Sect. V we analyze the potential implementations of future experiments. The main results of the work and some promising tasks for future activity are formulated in Sect. VI.

## II. STATEMENT OF THE PROBLEM AND CALCULATION TECHNIQUE

### A. Physical system and model

Let us consider one-dimensional (1D) structure of identical atoms placed over the line with period $a$ (see for example Fig.(1(a)). Some other types of artificial atom chains are shown in Figs. 2,3. Each atom is considered as a two-level Fermion system with transition frequency $\omega_0$. The location of the atoms in the lattice points is determined by the radius-vector $\mathbf{R}_p = \mathbf{e}pa$, $p$=0,1,…,$N$, $N$ is a number of atoms. We assume the tunneling to be predominant mechanism of interatomic coupling and neglect other ones (such as Förster and the radiation field transfer). As it was shown in [41], such assumption can be justified for a wide range of realistic parameter values.

Before starting the consideration, we will discuss the potential degree of correspondence of our model to the different types of chains. Three rather conventional simplifications have been done: i) we neglect all types of damping; ii) we use so called rotating-wave approximation (RWA) [26]; iii) we assume the atomic chain to be infinitely long and perfectly periodic.

Assumption i) relates to the case for scattering and radiation times strongly exceed the Bloch and Rabi periods. Because of RO for the quantum light are non-periodic, we imply here the value $t_R \cong \Omega_{\langle n \rangle}^{-1}$, as Rabi-period ($\langle n \rangle$ is the mean number of photons). These conditions are met with a large margin in superconductor junctions (artificial fluxonium atoms) [53-56]. The damping of BO and RO in heterostructures is defined by the electron scattering and decoherence, respectively. The recent progress in molecular beam epitaxy allowed achieving the values of scattering time $\tau \cong 10^{-10}$ s in the ultra-high-quality AlGaAs/GaAs heterostructures [57]. For this case BO may be considered as ballistic, while dephasing becomes the dominant component of the damping for coherent intersubband THz transitions (the typical values of characteristic times are 100-300 fs [50]). Such values are comparable with typical Bloch and Rabi periods; therefore the damping does not manifest itself. In [50] was experimentally observed the manifestation of RO for the pulse with duration 200 fs, which not exceeds the dephasing time and comparable with Rabi period (1-2 RO cycles over the pulse). For validity of our model to this case we will consider interaction with rather short pulse, whose duration, however, strongly exceeds the period of high-frequency filling and guaranties RWA validity.

The perfect periodicity means the identity and rather large number of atoms in the chain. Its implementation is rather simple for 2D-heterostructures (for example, the chain of 51 element was used in experiments [50]). The problem is not so easy for QD-chains because of QDs comprise hundreds or thousands of real atoms, with inevitable variations in size and shape and, consequently, unavoidable variability in their energies and relaxation times [58, 59]. One of the most promising technologies for applications in quantum photonics is the embedding of QDs within nanowires [59]. The QDs form at the apex of a GaAs/AlGaAs interface, are highly stable, and can be positioned with nanometer precision relative to the nanowire centre. As it was found [59], there is a chain of bright, nanoscale emitters in the red. QD-in-nanowire mimics very closely a two-level atom with high associated lifetime ($\cong 450$ ps). One more way is using a scanning tunneling microscope to create QD with identical, deterministic sizes [58]. As it was mentioned in Ref. 58, the reached digital fidelity opens the door to QD architectures free of intrinsic broadening. This makes reasonable to recommend our model of 1D QD-chain for visible light applications.

There exist different ways of theoretical description of complex quantum systems strongly interacted with quantum light. In quantum optics an "all-matter" picture is widely used, where the dynamics of light is integrated out (for example, in the optical Bloch equations [60,61]). In Refs [62,63] the light-matter interaction is treated in an "all-light" picture (Lippmann-Schwinger equation approach). We use as theoretical approach the probability amplitude method, generalized for the case of 1D chain driven via homogeneous dc field. It consists in solving of the Schrödinger equation for wave function, which is the superposition of various Wannier-Fock states for atom-light system.

To solve equations of motion for probability amplitudes analytically, we use the quasiclassical concept. The dressed electron is described as a wavepacket prepared with a well-defined quasimomentum. The motion of quasiparticle center of mass is governed by the Newton's law, while the internal degrees of freedom have been described by means of quantum theory based on the concept of electron-photon entanglement widely used in quantum optics [26,27]. Thus, the position of quasiparticle center periodically evolves with a Bloch frequency corresponding to the quasimomentum scanning a complete Brillouin zone. Such approach was used for BO of conductive electrons in [8,64] and will be adopted for the case of dressed electrons in this paper.

In general case, the equations of motion where integrated numerically with simplification through RWA. However, the use of the RWA may not describe the atom-light interaction when the coupling becomes sufficiently strong (ultrastrong coupling) [65]. Both lower and upper fundamental limitations of ultrastrong coupling of atoms and light were recently formulated in [66]. We bounded our consideration by the strong coupling regime assuming the lower limitation in [66] not to be reached.

### B. Hamiltonian in Wannier-Fock basis

` Let us denote $|a_p\rangle$ and $|b_p\rangle$ Wannier wave functions centered at *p*-th atom in the excited and ground states, respectively (Fig. 1). The two neighboring atoms are coupled via the electron tunneling, such that only intraband tunnel transitions are permitted. It means that the electron due to the tunneling can go from state $|a_p\rangle$ to state $|a_{p+1}\rangle$ and from state $|b_p\rangle$ to state $|b_{p+1}\rangle$ only. The transitions between ground and excited states of different atoms are forbidden: $\langle a_{p\pm 1}|b_p\rangle \approx 0$.

Let the chain be exposed to a single-mode quantum light, which electric field operator is $\hat{\mathbf{E}} = \hat{\mathbf{E}}^{(+)} + \hat{\mathbf{E}}^{(-)} = \mathbf{e}\sqrt{\hbar\omega/2\varepsilon_0 V}\left(\hat{a}+\hat{a}^+\right)$, where $V$ is the normalizing volume, $\mathbf{e}$ is the unit polarization vector, $\hat{\mathbf{E}}^{(+)} = \mathbf{e}\left(\hbar\omega/2\varepsilon_0 V\right)^{1/2}\hat{a}$, $\hat{\mathbf{E}}^{(-)} = \mathbf{e}\left(\hbar\omega/2\varepsilon_0 V\right)^{1/2}\hat{a}^+$ are positive and negative frequency components of field operator, $\hat{a},\hat{a}^+$ are creation-annihilation operators, respectively [an $\exp(-i\omega t)$ time dependence of the light is implicit]. The field assumed to be homogeneous over the chain axis. Such assumption corresponds to the chain excited by the normally incident wide laser beam, or specially-symmetric eigen-mode of microcavity, photonic crystal; etc. The chain is driven by electrostatic (dc) field directed along the axis with responsibility for BO. We will consider the case of dipole interaction in the regime of strong coupling and assume the resonant condition $\omega_0 \approx \omega$ to be fulfilled. The system under consideration exhibits complex single-particle BO of the electrons entangled with photons, for which the theoretical framework will be introduced.

The system is described by the total Hamiltonian (see Appendix A):

$$\hat{H} = \hat{H}_0 + \hat{H}_{ph} + \hat{H}_T + \hat{H}_{I,ph} + \hat{H}_{I,dc} \tag{1}$$

Here, the first term represents the free motion of the chain with absence of both light and dc field. It is given by $\hat{H}_0 = (\hbar\omega_0/2)\sum_p \hat{\sigma}_{zp}$, where $\hat{\sigma}_{zp} = |a_p\rangle\langle a_p| - |b_p\rangle\langle b_p|$. The second term $\hat{H}_{ph} = \hbar\omega\hat{a}^+\hat{a}$ is the

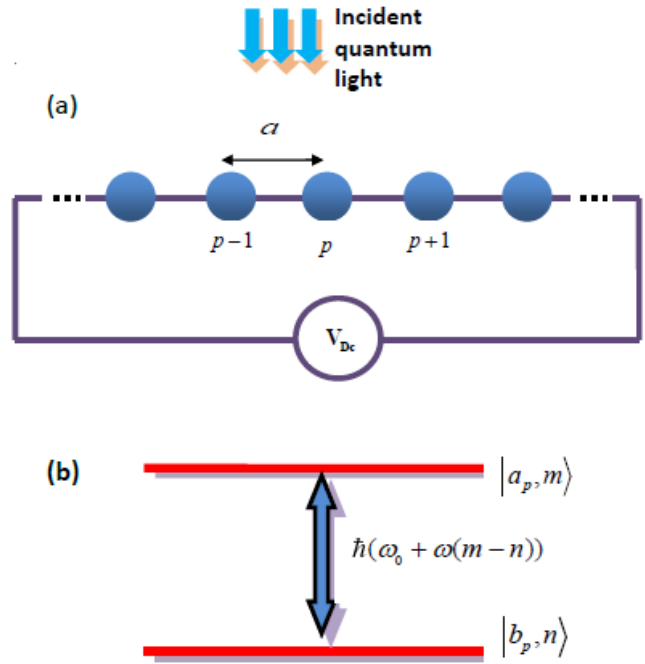

FIG. 1. (a) General illustration of the periodic two-level atomic chain used as a model indicating BO of electrons dressed with optical photons. It is excited with incident quantum light in the strong coupling regime. It is driven with dc voltage applied to the ends. The neighboring atoms are coupled via interatomic tunneling with different values of penetration for the ground and excited states. (b) Ground and excited energy levels of single two-level atom, separated by the transition energy $\hbar\omega_0$. The transition from the ground level to excited one (and vice versa) is accompanied with absorption (emission) of a single photon with energy $\hbar\omega$.

Hamiltonian of free electromagnetic field. The term

$$\hat{H}_T = -t_a \sum_p \left( |a_p\rangle\langle a_{p+1}| + |a_p\rangle\langle a_{p-1}| \right) - $$
$$- t_b \sum_p \left( |b_p\rangle\langle b_{p+1}| + |b_p\rangle\langle b_{p-1}| \right) \quad (2)$$

is Hamiltonian of electron tunneling, $t_{a,b}$ are the penetration energies of potential barrier at the excited and ground states, respectively.

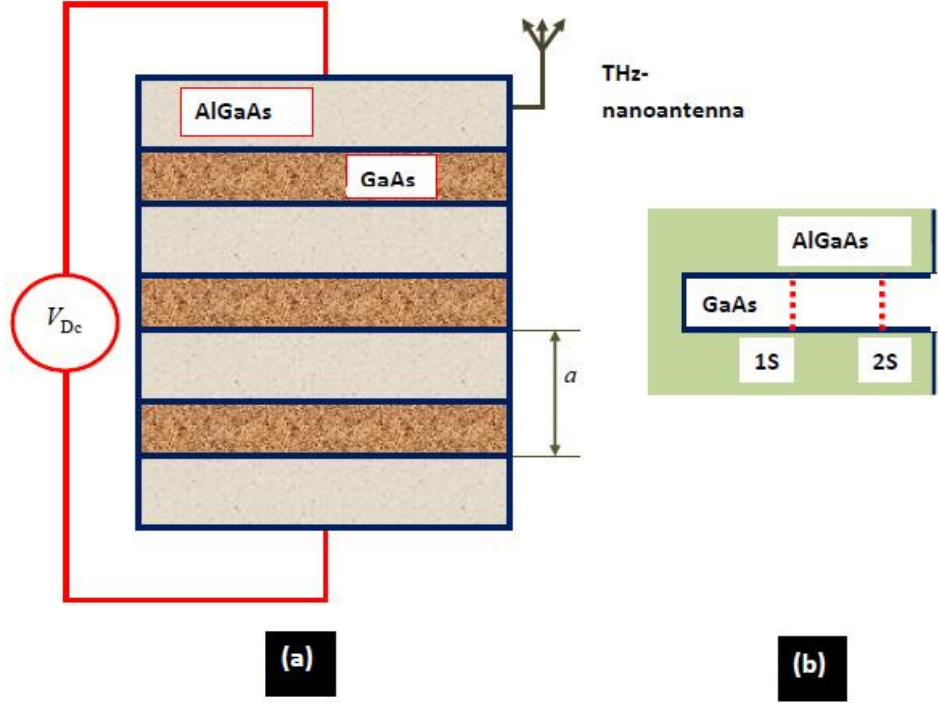

FIG 2. Schematic diagram of 2D semiconductor heterostructure. (a). *n*-type modulation-doped quantum well sample consisting from GaAs wells considered as an artificial atoms forming a chain. The atoms are separated from each other by AlGaAs barriers. The coupling is governed by the inter-barrier electron tunneling. The dc voltage applied to the ends of the system. The femtosecond THz pulse comes in through nanoantenna. (b) $1S \Leftrightarrow 2S$ intersubband transition, which is coherently excited by the pulse with narrow spectrum (central frequency is resonant with quantum transition).

The component

$$\hat{H}_{I,ph} = \hbar g \sum_p \hat{\sigma}_p^+ \hat{a} + \text{H.c.} \quad (3)$$

describes the atom-light interaction, where $g = -\sqrt{\omega/2\hbar\varepsilon_0 V}\,(\mathbf{e}\cdot\mathbf{d}_{ab})$ is the interaction constant, $\mathbf{d}_{ab}$ is the dipole moment, H.c. means Hermitian conjugation. The transition dipole moments in the chain assumed to be vectors of identical values and orientations. The operators $\hat{\sigma}_p^+ = |a_p\rangle\langle b_p|$, $\hat{\sigma}_p^- = |b_p\rangle\langle a_p|$ are creation-annihilation operators of excited state in the *p*-th atom. The Hamiltonian (3) is written in the RWA form [26].

The last term

$$\hat{H}_{I,dc} = e\sum_p (\mathbf{E}_{dc} \cdot \mathbf{R}_p)|a_p\rangle\langle a_p| + e\sum_p (\mathbf{E}_{dc} \cdot \mathbf{R}_p)|b_p\rangle\langle b_p| \quad (4)$$

describes the driving via dc field $\mathbf{E}_{dc}$ and is responsible for BO.

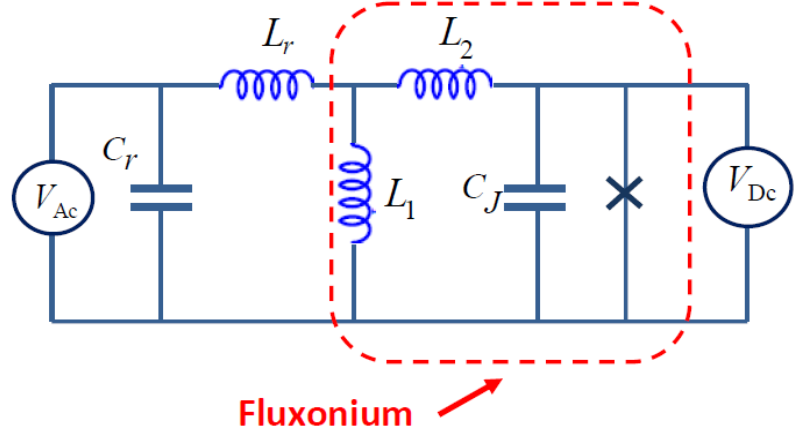

FIG. 3. Circuit diagram of a Josephson junction (crossed), which is inductively coupled with microwave resonator. The inductance $L_2$ describes the inductive environment of the junction. The inductance $L_1$ is a tool of control of the "junction – resonator" coupling (it is assumed $L_2 \gg L_1$). Josephson energy $E_J$ strongly exceeds charging energy $E_C$ ($E_J \gg E_C$). We model the resonator as a LC-circuit with effective inductance $L_r$ and effective capacitance $C_r$. The single fluxonium plays role of atomic chain due to the interaction with inductive environment [53-56] (BO take place in the space of quasi-charge).

### C. Equations of motion

The evolution of the system in the interaction picture is described by the Schrödinger equation $i\hbar\partial_t|\Psi\rangle = \hat{V}|\Psi\rangle$ with the interaction Hamiltonian given by

$$\hat{V} = \exp\left[\frac{i}{\hbar}(\hat{H}_0 + \hat{H}_{ph})\right](\hat{H}_{I,ph} + \hat{H}_{I,dc} + \hat{H}_T)\exp\left[-\frac{i}{\hbar}(\hat{H}_0 + \hat{H}_{ph})\right] \quad (5)$$

The state vector of the "atomic chain+light" system is

$$|\psi(t)\rangle = \sum_n \sum_p \{a_{p,n}(t)|a_p,n\rangle + b_{p,n}(t)|b_p,n\rangle\} \quad (6)$$

Here, $|b_p,n\rangle = |b_p\rangle \otimes |n\rangle$, $|a_p,n\rangle = |a_p\rangle \otimes |n\rangle$, are Fock states with $n$ photons and Wannier states centered around p-th atom in ground and excited state, respectively, $a_{p,n}$ and $b_{p,n}$ are the unknown probability amplitudes. From the Schrodinger equation we get following equations for the probability amplitudes:

$$i\hbar\frac{\partial a_{p,n}}{\partial t} = (\delta\varepsilon - \mathbf{E}_{dc} \cdot e\mathbf{R}_p)a_{p,n} + t_a a_{p+1,n} + t_a^* a_{p-1,n} - g\sqrt{n+1}\,b_{p,n+1}, \quad (7)$$

$$i\hbar\frac{\partial b_{p,n+1}}{\partial t} = -(\delta\varepsilon + \mathbf{E}_{dc} \cdot e\mathbf{R}_p)b_{p,n+1} + t_b b_{p+1,n+1} + t_b^* b_{p-1,n+1} - g\sqrt{n+1}\,a_{p,n} \quad (8)$$

where $\delta\varepsilon = (\varepsilon_a(0) - \varepsilon_b(0))/2$. In the limit $E_{dc} = 0$ the system (7), (8) goes to the corresponding equations obtained in [41] for Rabi-waves in QD chains.

### D. Studying observable values

We will study the values of two types. The first ones directly characterize the spatial-temporal behavior of electronic component and therefore are averaged over photonic state distribution. They are position-dependent, thus we will address to their densities per unit cell of the chain. From among these values we introduce:

i) *inversion density*, given by

$$W(p,t) = \sum_n \left( |a_{p,n}(t)|^2 - |b_{p,n}(t)|^2 \right) \tag{9}$$

and

ii) *tunneling current density* (see Appendix B)

$$\begin{aligned}J_{\text{Tunneling},p}(t) &= J^{(a)}_{\text{Tunneling},p}(t) + J^{(b)}_{\text{Tunneling},p}(t) = \\ &\quad -i\frac{e}{2}t_a \sum_n \left( a_{p-1,n}(t) - a_{p+1,n}(t) \right) a^*_{p,n}(t) \\ &\quad -i\frac{e}{2}t_b \sum_n \left( b_{p-1,n}(t) - b_{p+1,n}(t) \right) b^*_{p,n}(t) + c.c.\end{aligned} \tag{10}$$

The values of the second type are directly characterize the quantum-statistical properties of light and therefore are spatially averaged. From among these values we will use

iii) *photonic number distribution*

$$\tilde{p}(n,t) = \sum_p \left( |a_{p,n}(t)|^2 + |b_{p,n}(t)|^2 \right) \tag{11},$$

iv) *mean number of photons*

$$\langle n(t) \rangle = \sum_n n\tilde{p}(n,t) \tag{12}$$

where $\tilde{p}(n,t)$ is given by (11).

v) *photonic number variance*

$$\delta n(t) = \sum_n n^2 \tilde{p}(n,t) - \langle n(t) \rangle^2 \tag{13}$$

where $\langle n(t) \rangle$ is given by Eq. (12);

vi) *von Neumann quantum entropy*

$$S = \sum_j p_j \ln p_j \tag{14}$$

where $p_j$ are the weights of the various Fock states of the statistical light distribution [44].

## III. APPROXIMATE ANALYTICAL SOLUTION

### A. Preliminaries

The simplest but appearance model of BO is based on quasi-classical approximation. It results from Eqs. (7),(8) with $g = 0$. Electron-photon interaction disappears (subscript $n$ becomes unnecessary) and the equations become separated. The system reads

$$i\hbar \frac{\partial a_p}{\partial t} = -e(\mathbf{E}_{dc} \cdot \mathbf{R}_p) a_p + t_a a_{p+1} + t_a^* a_{p-1} \tag{15}$$

and the same for $b_p$. (the value $\delta\varepsilon$ becomes arbitrary and is taken $\delta\varepsilon = 0$). In the absence of dc field ($\mathbf{E}_{dc} = 0$) the eigenstate of the system corresponds to the electrons tunneling in the periodic potential and has a form of Bloch-wave $a_{p+1}(t) = a_p e^{i(\varphi - \nu t)}$, where $\varphi$ is a phase shift per unit cell, coupled with the eigen frequency, by $\nu = 2\hbar^{-1} t_a \cos\varphi$. The phase shift is coupled with the continuous quasi-momentum via relation $h = \varphi/a$ conventionally restricted to the first Brillouin zone $[-\pi/a; \pi/a]$. Under the influence of dc field, a given Bloch-state $|\Psi(h_0)\rangle = N^{-1/2} \sum_p |a_p\rangle e^{i(h_0 p a - \nu(h_0) t)}$ evolves up to a phase factor into the state $|\Psi(h(t))\rangle$ with $h(t)$ variation according to

$$\frac{dh}{dt} = -\frac{eE_{dc}}{\hbar}, \tag{16}$$

or $h(t) = -e\hbar^{-1} E_{dc} t + h_0$. Thus, this evolution is periodic with a Bloch frequency corresponding to the time required for the quasimomentum to scan a full Brillouin zone. It is described by the substitution $\varphi(t) = h(t) a = -\omega_B t + \varphi_0$, where $\omega_B = -e\hbar^{-1} E_{dc}$ is the Bloch frequency, $h_0 = \varphi_0 a$. The obtained solution relates also to the rather wide wave-packet with well-defined quasimomentum. The periodic motion of its geometrical center corresponds to the quasiclassical model of BO.

The main idea of the analytical solution considered next is based on hypothesis, that the total dynamics of quasiparticles described by Eqs. (8),(9) is reducible to the superposition of two interacting partial motions: i) internal motion, which is dictated by dressing; ii) external motion of the quasiparticle as a whole. The internal motion is of completely quantum origin, and doesn't allow any classical interpretation. It corresponds to Rabi-wave and plays role of the tunneling in the simplest case of BO considered before. The external motion is a motion of geometric center and will be described quasiclassically, following Eq. (16).

### B. Details

We consider the partial solutions of system (7)-(8) in the absence of dc field ($\mathbf{E}_{dc} = 0$) in the form $a_{p+1,n}(t) = a_{p,n}(t) e^{i\varphi}$, $b_{p+1,n}(t) = b_{p,n}(t) e^{i\varphi}$, $a_{p,n}(t) = \tilde{a}_{pn} e^{-i\nu t}$, $a_{p,n}(t) = \tilde{a}_{pn} e^{-i\nu t}$, $b_{p,n}(t) = \tilde{b}_{pn} e^{-i\nu t}$, where $\varphi$ is a given phase shift per unit cell, $\nu$ is the unknown eigen frequency, $\tilde{a}_{pn}, \tilde{b}_{pn}$ are unknown constant coefficients. Making use of Eqs. (8),(9) we obtain for them the matrix equation

$$\begin{pmatrix} \hbar\nu - 2t_a\cos\varphi & -g\sqrt{n+1} \\ -g\sqrt{n+1} & \hbar\nu - 2t_b\cos\varphi \end{pmatrix} \cdot \begin{pmatrix} \tilde{a}_{pn} \\ \tilde{b}_{p,n+1} \end{pmatrix} = 0 \qquad (17)$$

The eigenfrequencies are found from its characteristic equation

$$(\hbar\nu - 2t_a\cos\varphi)(\hbar\nu - 2t_b\cos\varphi) - g^2(n+1) = 0 \qquad (18)$$

and given by

$$\hbar\nu_{1,2}(\varphi) = \left\{(t_a + t_b)\cos\varphi \pm \sqrt{(t_a - t_b)^2\cos^2\varphi + g_n^2}\right\}, \qquad (19)$$

where $g_n = g\sqrt{n+1}$. The correspondent eigenstates are

$$|\Psi_{n,1}\rangle = \frac{1}{\sqrt{N(1+\Delta_n^2(\varphi))}} \sum_p e^{ip\varphi}\left(|a_p,n\rangle e^{-i\left(\frac{\omega}{2}+\nu_1\right)t} - \Delta_n(\varphi)|b_p,n+1\rangle e^{i\left(\frac{\omega}{2}-\nu_1\right)t}\right) \qquad (20)$$

$$|\Psi_{n,2}\rangle = \frac{1}{\sqrt{N(1+\Delta_n^2(\varphi))}} \sum_p e^{ip\varphi}\left(\Delta_n(\varphi)|a_p,n\rangle e^{-i\left(\frac{\omega}{2}+\nu_2\right)t} + |b_p,n+1\rangle e^{i\left(\frac{\omega}{2}-\nu_2\right)t}\right) \qquad (21)$$

where

$$\Delta_n(\varphi) = \frac{g_n}{(t_a - t_b)\cos\varphi + \sqrt{(t_a - t_b)^2\cos^2\varphi + g_n^2}} \qquad (22)$$

The states (14)-(15) describe the travelling of transitions between the states $|a_p,n\rangle$ and $|b_p,n+1\rangle$ along the chain (so called Rabi-waves [41]).

The periodicity of the lattice leads to a band structure of the energy spectrum of the dressed electrons in the entangled eigenstates $|\Psi_{\alpha,n}(h)\rangle$ with the corresponding eigenenergies $\hbar\nu_{1,2}(h,n)$. They are labeled by the discrete number of photons $n$ and the continuous quasimomentum $h$. Under the influence of dc field, weak enough not to induce interband transitions, the state $|\Psi_{\alpha,n}(h_0)\rangle$ evolves up to a phase factor into the state $|\Psi_{\alpha,n}(h(t))\rangle$ with $h(t)$ variation according to Eq. (16). This evolution is periodic with a Bloch frequency $\omega_B$ corresponding to the time required for the quasimomentum to scan a full Brillouin zone. It leads to the exchange

$$\Delta_n(\varphi) \to \Delta_n(t) = \frac{g_n}{(t_a - t_b)\cos(\omega_B t - \varphi_0) + \sqrt{(t_a - t_b)^2\cos^2(\omega_B t - \varphi_0) + g_n^2}} \qquad (23)$$

The distribution of photonic probabilities for the state $|\Psi_{n,1}\rangle$ is given by

$$p_1(m,t) = \begin{cases} \dfrac{1}{1+\Delta_n^2(t)}, & m=n \\ \dfrac{\Delta_n^2(t)}{1+\Delta_n^2(t)}, & m=n+1 \\ 0, & \text{others} \end{cases} \qquad (24)$$

The average number of photons is

$$\langle n_1(t) \rangle \approx n + \frac{\Delta_n^2(t)}{1+\Delta_n^2(t)}, \qquad (25)$$

where the value $\Delta_n(t)$ is given by relation (23). The similar elementary calculations for the state $|\Psi_{n,2}\rangle$ give

$$\langle n_2(t) \rangle \approx n + \frac{1}{1+\Delta_n^2(t)} \qquad (26)$$

For photonic number variances we have

$$\delta n_1 = \delta n_2 = \frac{\Delta_n^2(t)}{\left(1+\Delta_n^2(t)\right)^2} \qquad (27)$$

Here, von Neumann entropy [44] accumulated at the given states of light is equal for both states and given by

$$S_{n,1}(t) = S_{n,2}(t) = \frac{1}{1+\Delta_n^2(t)} \ln\left(\frac{1}{1+\Delta_n^2(t)}\right) + \frac{\Delta_n^2(t)}{1+\Delta_n^2(t)} \ln\left(\frac{\Delta_n^2(t)}{1+\Delta_n^2(t)}\right) \qquad (28)$$

Due to BO, such characteristics of quantum light as the photonic probabilities, the average number of photons and the Neiman entropy are oscillatory functions with Bloch frequency $\omega_B$. The amplitude of these non-monochromatic oscillations is strongly dependent on the relation between penetrations of potential barrier in the ground and excited states. In particular, as obey from Eq. (17), for the case of identical tunneling penetrations at the ground and excited states ($t_a \to t_b$) the value $\Delta_n$ becomes constant, and oscillations are vanish. The oscillating effect is suppressed likewise with the increasing of photons number $n$. As a consequence, a wave packet of entangled electron-photon prepared with a well-defined quasimomentum will also oscillate in position with an amplitude $2\hbar(t_a + t_b)/eE_{dc}$.

Finally, we note that there is an interesting ability of quantum state control via adiabatically turning the dc field on and off. Starting from $E_{dc} = 0$ we proceed to turn on $E_{dc}$ until its maximal value. Thereafter we slowly turn $E_{dc}$ off and end it at the moment $t = T$. This process corresponds to the turning the phase from initial value $\varphi = \varphi_0$ until the final one given by

$$\Phi = \varphi(T) = \varphi_0 - e\hbar^{-1} a^{-1} \int_0^T E_{dc}(\tau) d\tau \qquad (29)$$

The optimal choose of the turning time $T$ makes the arbitrary value of the phase turn to be reachable. As a result, the dc field via BO becomes an effective tool of control of quantum light statistics. In particular, the photonic probabilities, degree of electron-photon entanglement and Neumann entropy may be adiabatically changed from minimal to maximal values (and vice versa) via adiabatic turning of dc field.

# IV. NUMERICAL MODELLING AND DISCUSSION

In this section we will show and discuss the numerical results for the different types of initial states. The system (8)-(9) was solved with Crank-Nicolson numerical integration technique [67]. We has used periodic Born-Von Karman relations $a_{1,n} = a_{N,n}, b_{1,n} = b_{N,n}$ as boundary conditions [41]. Let us note that concrete form of boundary conditions has no physical importance in our case because the area of oscillations is placed rather distantly from the ends of the chain. For simplicity, we will limit our consideration by the case of zero detuning ($\omega = \omega_0$).

## A. Coherent photonic initial state

Let us assume the electron and photon subsystems to be initially non-interacting. The electronic subsystem is prepared as a coherent superposition of Gaussian wave packets

$$a_p(0) = a_0 e^{-\frac{(p-u_a)^2 a^2}{\sigma_a^2}} e^{ik_a pa} \qquad (30)$$

$$b_p(0) = b_0 e^{-\frac{(p-u_b)^2 a^2}{\sigma_b^2}} e^{ik_b pa} \qquad (31)$$

where $a_0, b_0$ are arbitrary complex values satisfy the normalization condition, $u_{a,b}, \sigma_{a,b}, k_{a,b}$ are the position of Gaussian center, effective Gaussian width and the initial value of quasimomentum at the ground and excited states, respectively. The photonic subsystem is prepared in the coherent state given by Poisson distribution $p(n) = \langle n \rangle^{n/2} e^{-\langle n \rangle/2} / \sqrt{n!}$, where $\langle n \rangle$ is the mean number of photons. The total wavefunction at the moment $t=0$ is given by

$$|\psi(0)\rangle = \sum_n \frac{\langle n \rangle^{n/2}}{\sqrt{n!}} e^{-\frac{\langle n \rangle}{2}} \sum_p \left( a_0 e^{-\frac{(p-u_a)^2 a^2}{\sigma_a^2}} e^{ik_a pa} |a_p, n\rangle + b_0 e^{-\frac{(p-u_b)^2 a^2}{\sigma_b^2}} e^{ik_b pa} |b_p, n\rangle \right) . \qquad (32)$$

In Fig. 4 we plotted the temporal dynamics of inversion for existence and absence of driven dc field. One can see that inversion behavior for these two cases is rather different. The case of zero dc field (Fig. 4(a)) corresponds to the well known scenario of collapse-revivals given by Jaynes-Cummings model for the single atom [26]. The initial spatial distribution is not disturbed over the motion via inter-atomic tunneling. For the case of dc field existence the collapse-revivals picture drifts due to BO: the collapses and revivals appear at the different segments of the atomic chain.

In Fig. 5 we plotted the time dependence of the tunneling current. For the absence of light (Fig 5(a)) the current exhibits the periodic dynamics with Bloch frequency. This picture is dramatically changed due to the light-chain coupling (Fig 5(b)). The tunnel current is modulated in exact compliance with the collapse-revivals picture. This result is rather counterintuitive: it demonstrates the influence of the quantum-light statistics on the low-frequency motion of the charges driven separately by dc field. It makes the spectra of the tunneling current more wide and various wherewith in the case of ordinary BO. It opens a new promising ways in the spectroscopy of nano-circuits and nano-devices based on the synthesis of quantum optical and dc tools of control [51, 52].

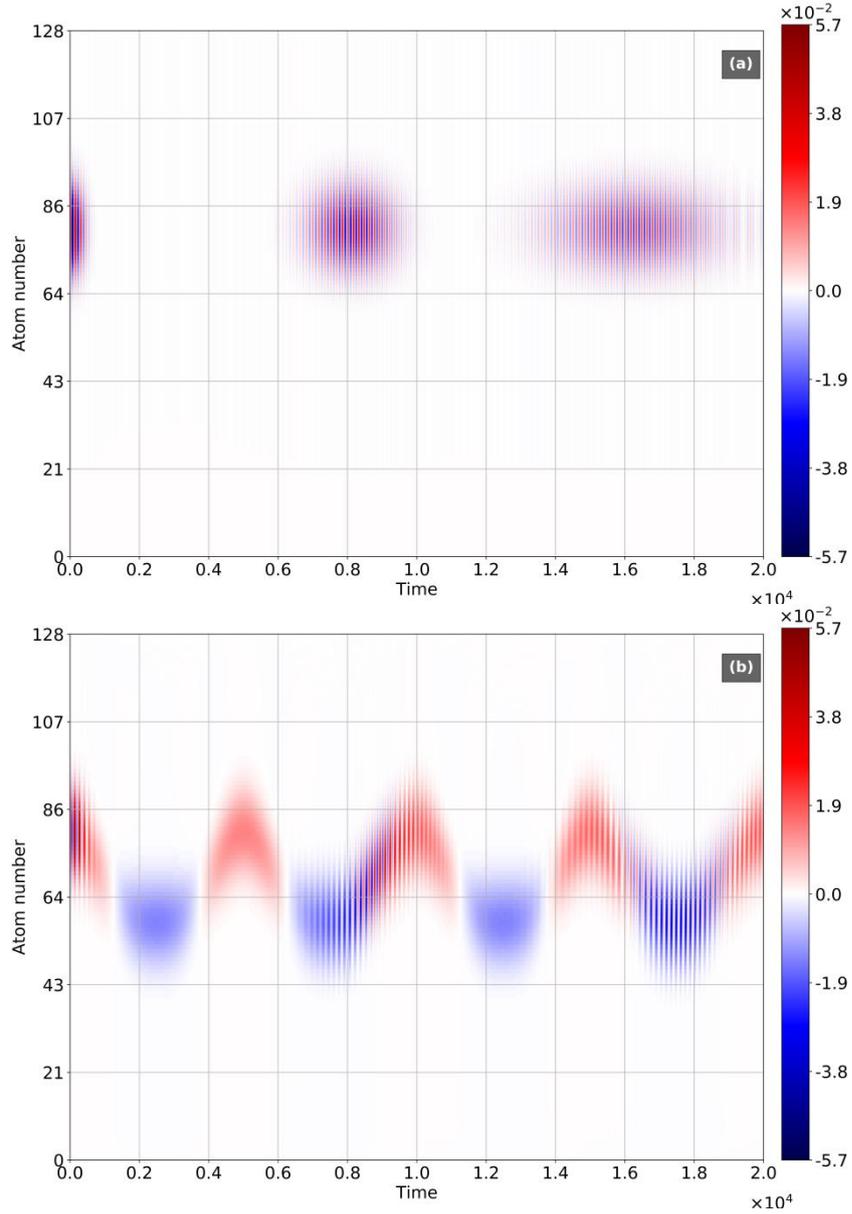

FIG. 4 Space-time distribution of inversion density for light in the coherent state with $<n>=25$. All frequencies are normalized to the frequency of quantum transition $\omega_0$. The number of atoms $N=128$, Rabi-frequency $\Omega_{\langle n \rangle} = 0.025$, electron initial Gaussian with $a_0 = 1, b_0 = 0$, $u_a = 80, \sigma_a = 10, k_a = 0$. (a) $\omega_B = 0$, $t_a/\omega_B = t_b/\omega_B = 10$. (b) $\omega_B = 0.0008$, $t_a/\omega_B = 10, t_b/\omega_B = 1$.

In Fig 6 we plot the dynamics of mean number of photons. Again, we see the mutual influence of RO and BO. For dc field absence (Fig. 6(b)) the collapse-revivals scenario corresponds the Jaynes-Cummings model for single atom [26]. It represents the quantum interference of the spectral components with different Rabi-frequencies $\Omega_n$ and different relative amplitudes [26]. For dc field appearance (Fig. 6(a)) has been appeared one more type of interference components with frequencies $m\omega_B$ ($m$ is integer value). Among them, the term with $m=1$ is dominant, and its amplitude is comparable with RO components. As a result, the mean number of photons is modulated with Bloch frequency $\omega_B$. Therefore, the dc field via BO opens one more way of engineering of the quantum light states. This effect depends on the large number of physical factors, such as dc field value, energy and dipole moment of quantum transition for the atom, tunnel coupling, etc. Concluding, the results of numerical modeling agree with the simple analytical model, developed in Sect. III. They are promising for different applications in quantum computing, quantum informatics, etc.

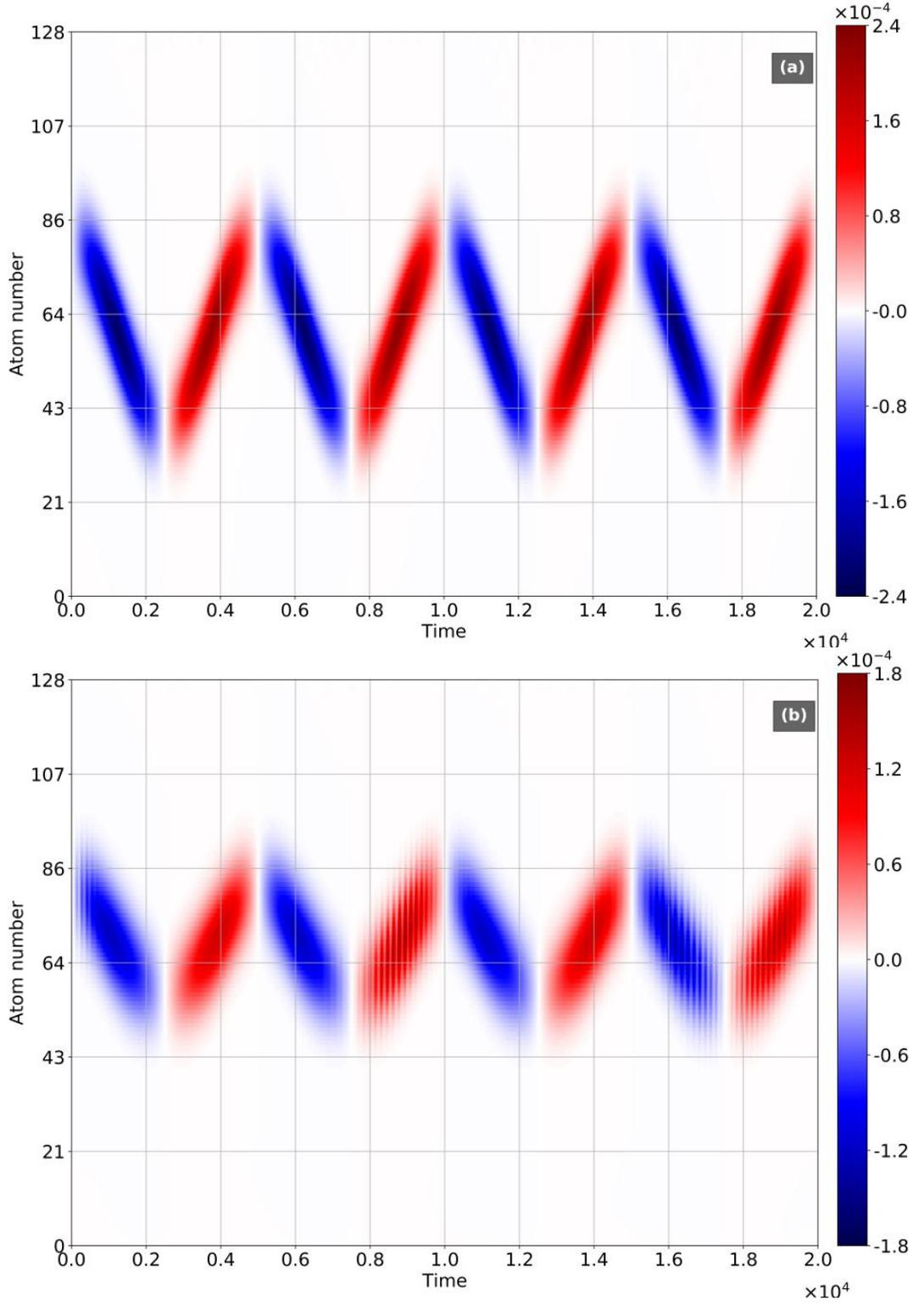

FIG. 5. Space-time distribution of tunnel current density for light in the coherent state with $<n>=25$. All frequencies are normalized to the frequency of quantum transition $\omega_0$. The number of atoms $N=128$, Rabi-frequency $\Omega_{\langle n \rangle} = 0.025$, $\omega_B = 0.0008$, electron initial Gaussian with $a_0 = 1, b_0 = 0$, $u_a = 80, \sigma_a = 10, k_a = 0$. (a) $\omega_B = 0$, $t_a/\omega_B = t_b/\omega_B = 10$. (b) $\omega_B = 0.0008$, $t_a/\omega_B = 10, t_b/\omega_B = 1$.

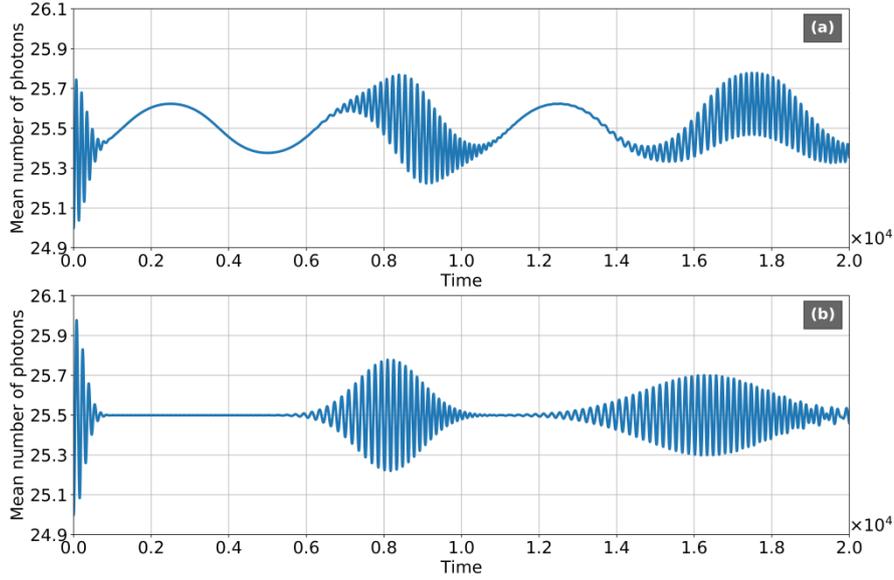

FIG. 6. Mean number of photons for light in the coherent state with $<n>=25$. All frequencies are normalized to the frequency of quantum transition $\omega_0$, $\Omega_{\langle n \rangle}=0.025$. (a) The chain with the number of atoms $N=128$, $\omega_B=0.0008$, electron initial Gaussian with $a_0=1, b_0=0$, $u_a=80, \sigma_a=10, k_a=0$, $t_a/\omega_B=10$, $t_b/\omega_B=1$. (b) Dressed single atom (Jaynes-Cummings model).

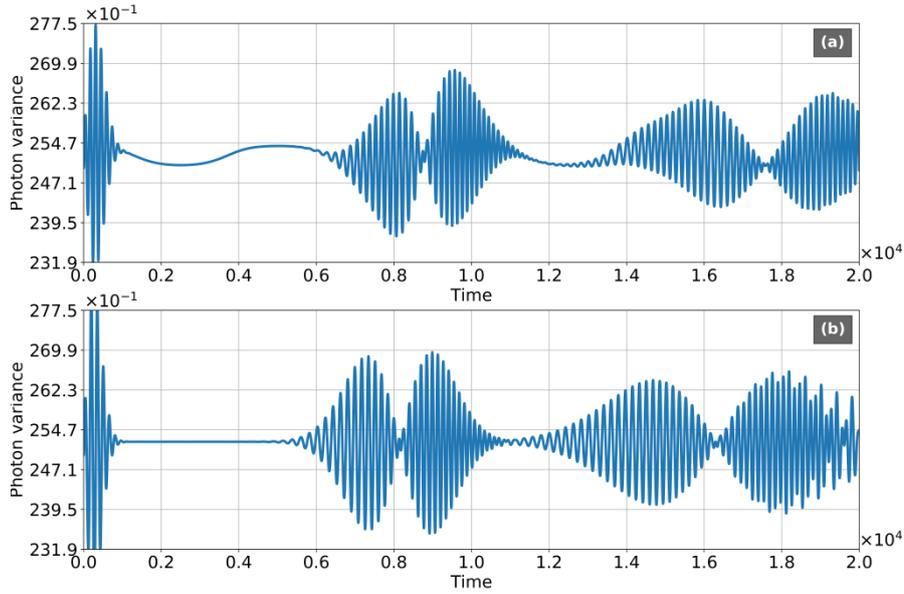

FIG. 7. Photon number variance for light in the coherent state with $<n>=25$. All frequencies are normalized to the frequency of quantum transition $\omega_0$, $\Omega_{\langle n \rangle}=0.025$. (a) The chain with the number of atoms $N=128$, $\omega_B=0.0008$, electron initial Gaussian with $a_0=1, b_0=0$, $u_a=80, \sigma_a=10, k_a=0$, $t_a/\omega_B=10$, $t_b/\omega_B=1$. (b) Dressed single atom (Jaynes-Cummings model).

In Fig 7 we plot the dynamics of photon number variance. It exhibits again the collapse-revivals scenario, however, the every RO revival splits on two parts. In contrast with the single atom (Fig. 7b), the variance in our case is modulated with Bloch-frequency in the surrounding of value $\delta n = \langle n \rangle$ even in the collapse areas. The modulation amplitude is rather small. It is not surprising, if one accounts that coherent state with the large number of photons is characterized by minimally possible non-classicality. In another words, it gives the classical limit of quantum light, while the classical light doesn't experience the inverse influence of BO [51, 52].

## B. Vacuum photonic initial state

Now we will consider the initial state of vacuum for light and coherent superposition of Gaussians for atomic chain. It is given by the wave function

$$|\psi(0)\rangle = \sum_p \left( a_0 e^{-\frac{(p-u_a)^2 a^2}{\sigma_a^2}} e^{ik_a pa} |a_p, 0\rangle + b_0 e^{-\frac{(p-u_b)^2 a^2}{\sigma_b^2}} e^{ik_b pa} |b_p, 0\rangle \right) \quad (33)$$

where $a_0, b_0$ are the arbitrary coefficients satisfying normalization condition.

In Fig. 8 we plot the evolution of inversion. Here, BO leads to the separation of areas of positive and negative values of inversion. The particle in the excited state (positive inversion) exhibits the single oscillation of rather high amplitude in the course a half of Bloch circle. During the other part of Bloch circle the particle is in rest, which means its appearance in the ground state (negative inversion). The reason for this is the weak value of transparence of potential barrier at the ground level used in the shown calculations. In contrast with single atom scenario, vacuum RO between maximal and minimal values of inversion occur with Bloch frequency $\omega_B$ instead of Rabi frequency $\Omega_0 = 2g$. The tunnel current induced by BO is dramatically changed due to the interaction with light (Fig 9). It periodically evolves through one pulse after another with frequency $\omega_B$ instead of continuous periodic behavior in the case of ordinary BO [8-25]. Such evolution is accompanied by the modulation with Rabi frequency. The shown results demonstrate high and various mutual interactions of RO and BO.

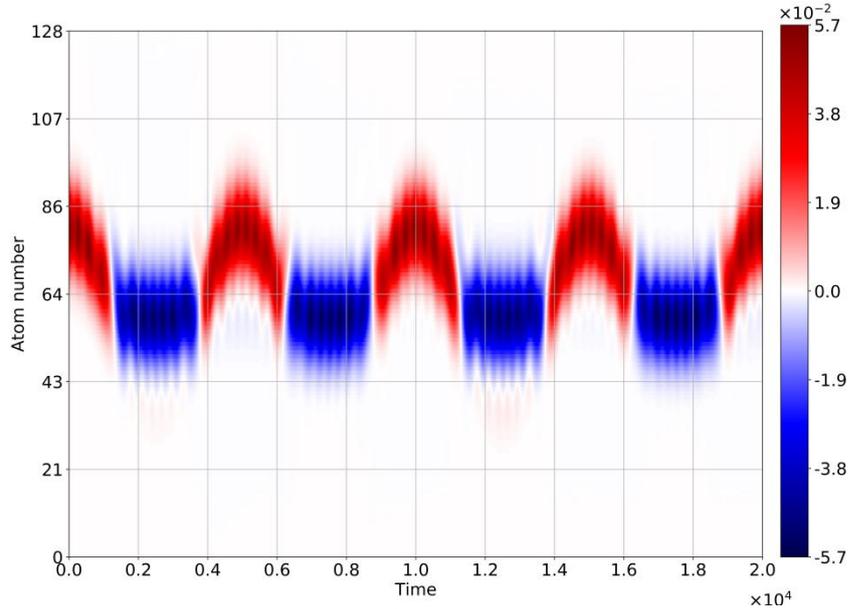

FIG. 8 Space-time distribution of inversion density for light in vacuum state. All frequencies are normalized to the frequency of quantum transition $\omega_0$. The number of atoms N=128, Rabi-frequency $\Omega_0 = 0.025$, $\omega_B = 0.0008$ electron initial Gaussian with $a_0 = 1, b_0 = 0$, $u_a = 80, \sigma_a = 10, k_a = 0$, $t_a/\omega_B = 10, t_b/\omega_B = 1$.

Figure 10 shows the value of mean number of photons. Its temporal behavior may be rather simply understood from the quasi-classical point of view developed in Sect. III. The processes of photons emission and absorption correspond to the rapid jumps from value $\langle n \rangle \approx 0.1$ to $\langle n \rangle \approx 0.9$ (and vice versa). Comparing Fig. 10 with Fig. 8, one can see these jumps occurrence in a small vicinity of the stops and turns of the wavepacket. The neighboring stoping points are separated in time by the half of a Bloch circle. Sufficiently far from a stopping point the motion is characterized by rather small acceleration. It means the existence of rather weak fluctuations of photonic number, dictated by RO and taking place with Rabi-frequency $\Omega_0 = 2g$. This scenario dramatically differs from RO in the single atom (Fig. 10b), where the acts of photons emission-absorption take place with Rabi-frequency. Note again, that such physical behavior agrees with evolution of the inversion (Fig. 8).

Figure 11a displays the evolution of photonic variance. As it was mentioned above, the electrons and photons in the initial state are uncorrelated. Thus, electron-photon entanglement is poor at the intervals of motion between the stopping points, whereby the variance oscillates only due to photon number fluctuations of rather small amplitude with vacuum Rabi-frequency. The processes of photons emission-absorption are accompanied the strong narrow peaks of variance. Thus, the vicinities of stopping points are characterized by the strong non-classicality of light and high degree of its entanglement with electronic sub-system. This scenario strongly differs from the case of the single atom (Fig.11b) with harmonic oscillations of variance with Rabi-frequency. The obtained results show, that such manifestations of non-classicality of light as electron-photon entanglement and accumulated entropy become controllable by dc field via BO. As it is well known [26], the physical origin of vacuum RO is a manifestation of spontaneous emission of the single exited state in strong coupling regime. The obtained dynamics qualitatively support the conclusion of early paper [68], that BO dramatically changes the physical picture of spontaneous emission in the atomic chains.

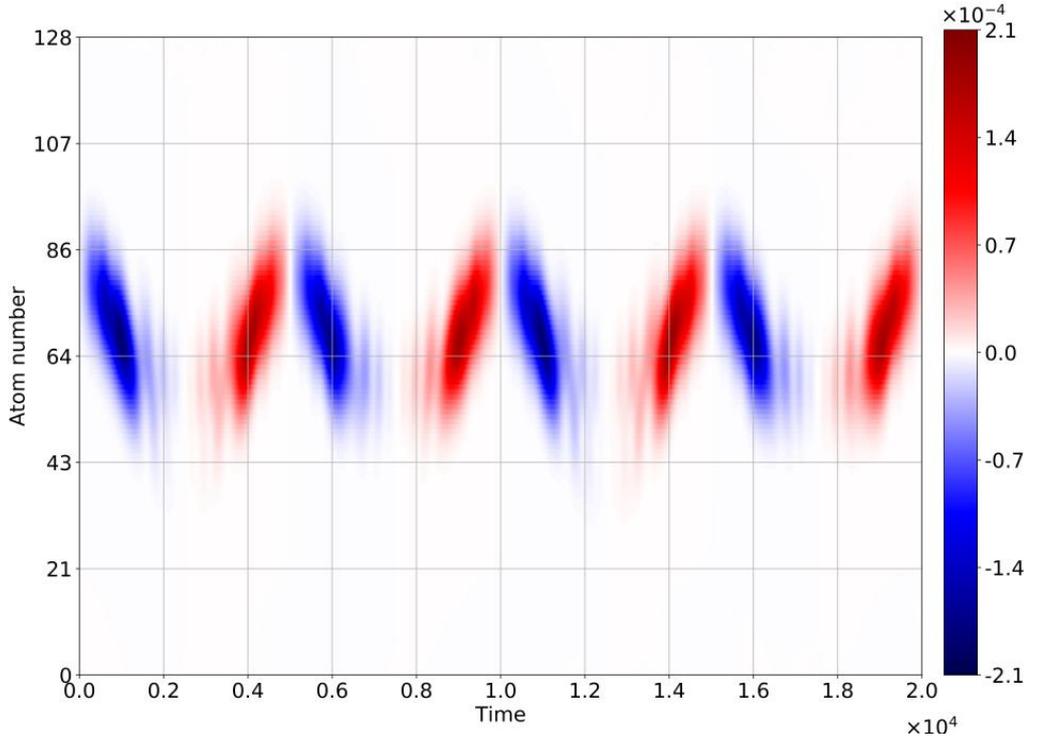

FIG. 9. Space-time distribution of tunnel current for light in vacuum state. All frequencies are normalized to the frequency of quantum transition $\omega_0$. The number of atoms $N$=128, Rabi-frequency $\Omega_0 = 0.025$, $\omega_B = 0.0008$ electron initial Gaussian with $a_0 = 1, b_0 = 0$, $u_a = 80, \sigma_a = 10, k_a = 0$, $t_a/\omega_B = 10, t_b/\omega_B = 1$.

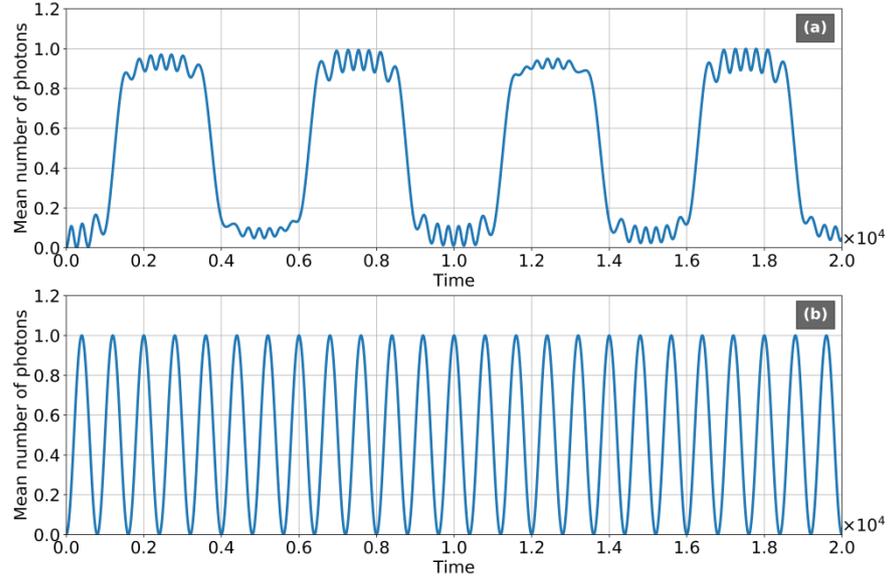

FIG. 10. Mean number of photons for light in vacuum state. All frequencies are normalized to the frequency of quantum transition $\omega_0$, Rabi-frequency $\Omega_0 = 0.025$, (a) The chain with the number of atoms $N=128$, $\omega_B = 0.0008$, electron initial Gaussian with $a_0 = 1, b_0 = 0$, $u_a = 80, \sigma_a = 10, k_a = 0$, $t_a/\omega_B = 10$, $t_b/\omega_B = 1$. (b) Dressed single atom (Jaynes-Cummings model).

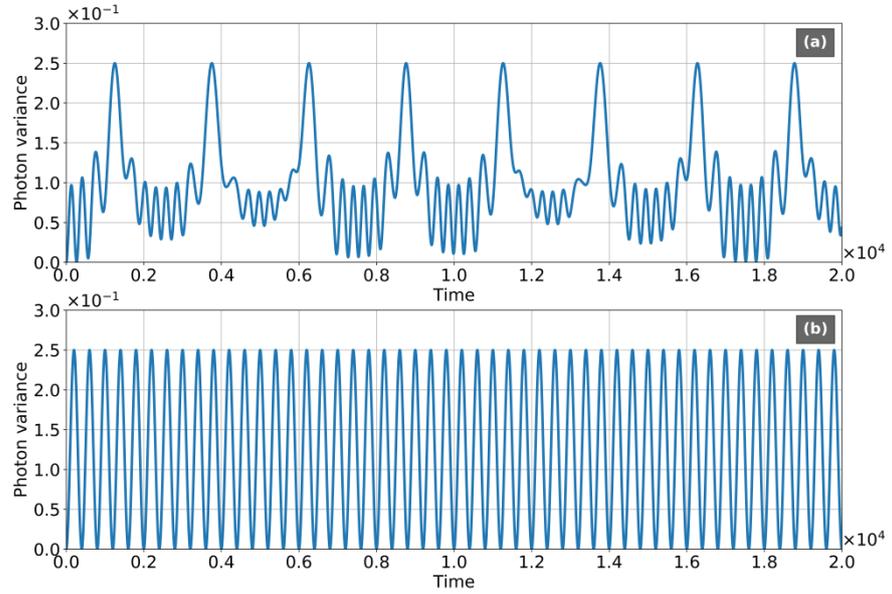

FIG. 11. Number of photon variance for light in vacuum state. All frequencies are normalized to the frequency of quantum transition $\omega_0$, Rabi-frequency $\Omega_0 = 0.025$, (a) The chain with the number of atoms $N=128$, $\omega_B = 0.0008$, electron initial Gaussian with $a_0 = 1, b_0 = 0$, $u_a = 80, \sigma_a = 10, k_a = 0$, $t_a/\omega_B = 10$, $t_b/\omega_B = 1$. (b) Dressed single atom (Jaynes-Cummings model).

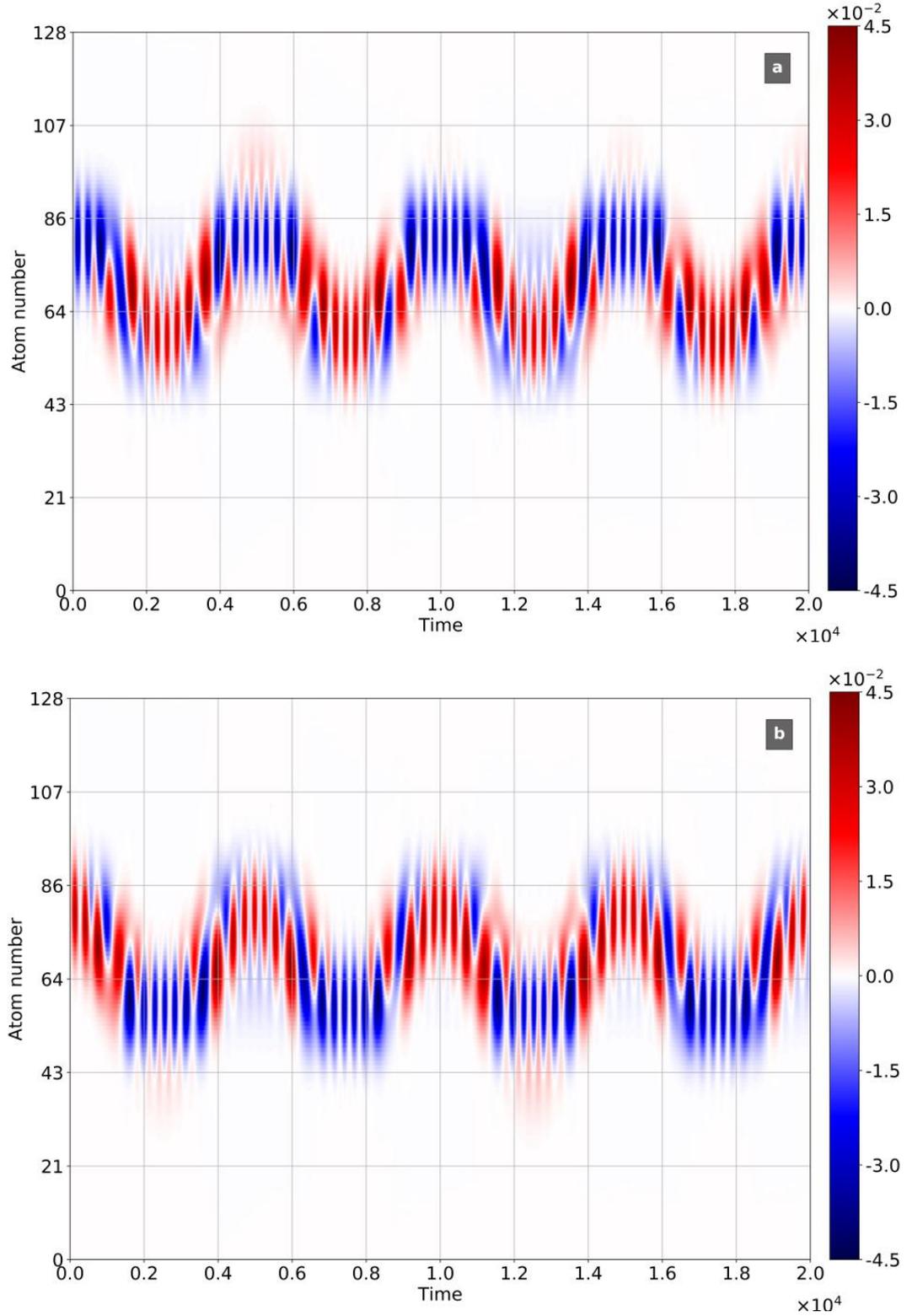

FIG. 12. Space-time distribution of inversion density for the case of atom-light initial entanglement. All frequencies are normalized to the frequency of quantum transition $\omega_0$. The number of atoms $N=128$, Rabi-frequency $\Omega_0 = 0.025$, $\omega_B = 0.0008$, electron initial Gaussian with $a_0 = 1, b_0 = 0$, $u_a = 80, \sigma_a = 10, k_a = 0$, $t_a/\omega_B = 10, t_b/\omega_B = 1$. (a) In-phase initial condition. (b) Out-of-phase initial condition.

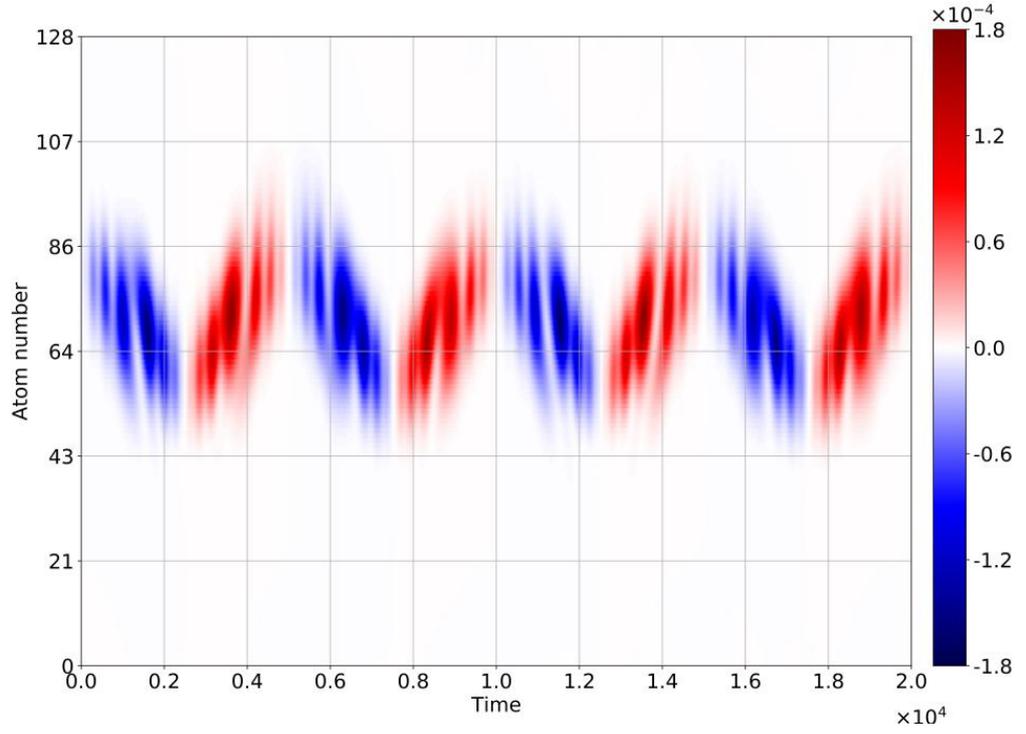

FIG. 13. Space-time distribution of tunnel current density for the case of atom-light initial entanglement. In-phase initial condition. All frequencies are normalized to the frequency of quantum transition $\omega_0$. The number of atoms $N=128$, Rabi-frequency $\Omega_0 = 0.025$, $\omega_B = 0.0008$, electron initial Gaussian with $a_0 = 1, b_0 = 0$, $u_a = 80, \sigma_a = 10, k_a = 0$, $t_a/\omega_B = 10, t_b/\omega_B = 1$.

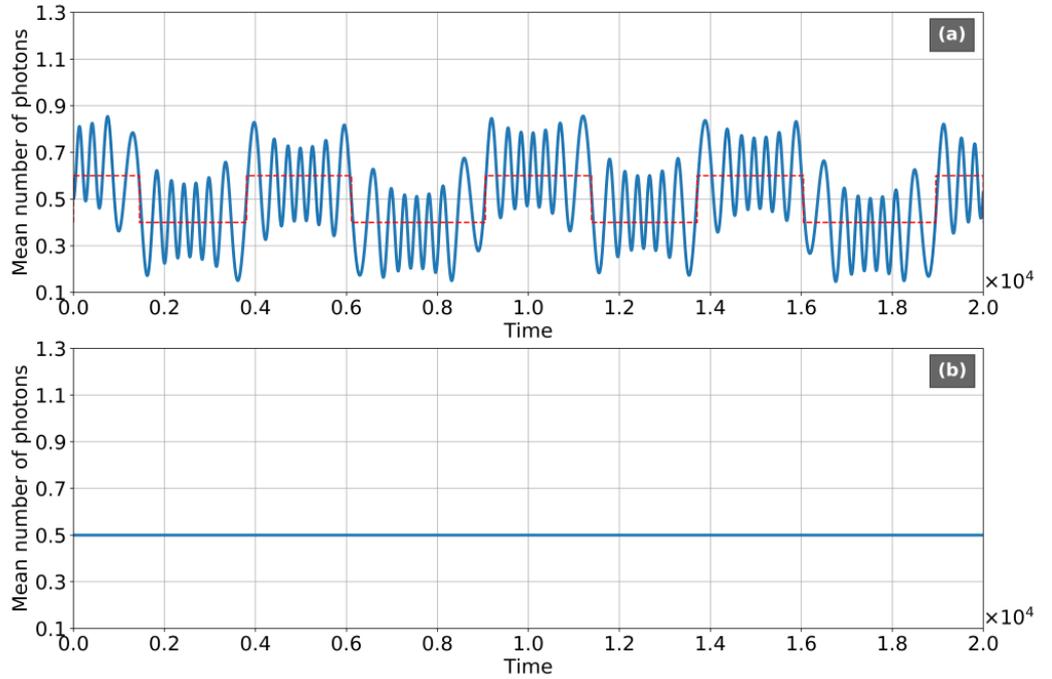

FIG. 14. Mean number of photons for the case of atom-light initial entanglement. In-phase initial condition. All frequencies are normalized to the frequency of quantum transition $\omega_0$. Rabi-frequency $\Omega_0 = 0.025$, (a) The chain with number of atoms $N=128$, $\omega_B = 0.0008$, electron initial Gaussian with $a_0 = 1, b_0 = 0$, $u_a = 80, \sigma_a = 10, k_a = 0$, $t_a/\omega_B = 10, t_b/\omega_B = 1$. (b) Dressed single atom (Jaynes-Cummings model).

## C. Entangled photon-electron initial state

Here, we consider the case of electron-photon entanglement prepared initially before the the driving fields would be switching on. It will be considered in-phase initial condition, which reads

$$|\psi_n(0)\rangle = C_0 \sum_p e^{-\frac{(p-u)^2 a^2}{\sigma^2}} \left(|a_p, n\rangle + |b_p, n+1\rangle\right) \tag{34}$$

or out-of-phase initial condition

$$|\psi_n(0)\rangle = C_0 \sum_p e^{-\frac{(p-u)^2 a^2}{\sigma^2}} \left(|a_p, n\rangle - |b_p, n+1\rangle\right) \tag{35}$$

where $C_0$ is normalization constant. The value of inversion for the states (34), (35) is equal to zero.

To illustrate the mutual influence of high-frequency and low-frequency motions, we show the spatial-temporal evolution of inversion for initial conditions (34), (35) (Figure 12). One finds out of the phase oscillations of inversion with Bloch frequency. Two halves of every Bloch period corresponds to the excited and ground state, respectively. For initial state (34) the system starts to evolve in the direction of the ground state (negative inversion) (see Fig. 12(a)), for initial state (35) it begins the motion in the opposite direction. (see Fig. 12(b)). Here, BO of inversion is modulated with RO. Such dynamics is similar to the shown for the vacuum initial state at Fig. 8, however, the modulation depth strongly increases due to the influence of initial entanglement (the inversion evolves over the single Rabi-circle between zero and $\pm 1$). In Fig. 13, we show the tunnel current for initial state (34), which as well as inversion exhibits the high level of RO modulation. Comparing it with Fig.9, one finds again the strong influence of initial electron-photon correlations.

Figure 14(a) displays the evolution of mean number of photons for the state (34). It may be considered as a periodic system of step-like beatings (dashed line), modulated with high-frequency oscillations (solid curve). The value $\langle n \rangle$ averaged with respect to the step-period is approximately equal to 0.5, which is equal to the main photon number for both states $|\Psi_{n1,2}\rangle$ given by Eqs (20),(21). The general physical interpretation of Fig. 14(a) may be done similar Sect. IV B from the quasi-classical point of view as an interference of this states (of course, we speak here about moving wavepackets, but not about perfect travelling waves). Again, the photon emission-absorption with Bloch-period takes place in the vicinity of stopping points. One can see comparing Fig. 14(a) with Fig 12(a), that the moments of shut-down are agreed with zeros of inversion. Translatory motion of the wavepacket is accompanied by RO between eigen-states, which correspond to high-frequency fluctuations with Rabi-frequency. In Fig.15(a) we plot the dynamics of photonic variance. In contrast with vacuum photonic state, the variance is maximal at the intervals of translatory motion and strongly decreases in the vicinities of stopping points. The reason for it is that translatory motion corresponds to the statement in one of the eigen states (20),(21). Everyone of this states is characterized by the maximal level of electron-photon entanglement. Photons emission-absorption means the mutual transformation of the states (20), (21). , It leads to creation the coherent superposition of these states for a short time with comparable probability amplitudes. It leads to the breaking of electron-photon correlations. Again, RO-fluctuations are imposed to this ideal picture similar to the case of vacuum photonic state. The photonic variance in this oscillations reaches the value $\delta n = 0.25$, which corresponds to the maximal one for two states (20),(21).

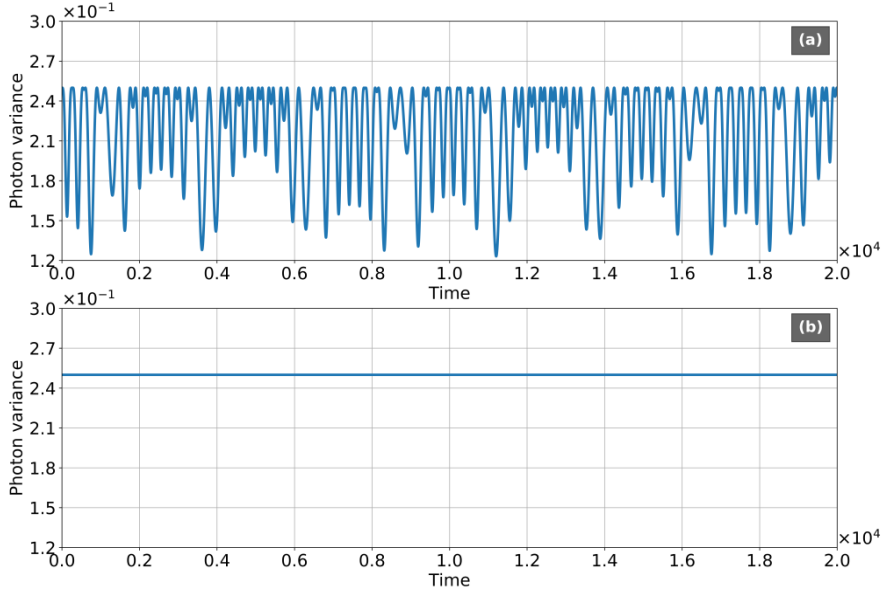

FIG. 15. Photon number variance for the case of atom-light initial entanglement. In-phase initial condition. All frequencies are normalized to the frequency of quantum transition $\omega_0$. Rabi-frequency $\Omega_0 = 0.025$, (a) The chain with number of atoms $N=128$, $\omega_B = 0.0008$, electron initial Gaussian with $a_0 = 1, b_0 = 0$, $u_a = 80, \sigma_a = 10, k_a = 0$, $t_a/\omega_B = 10, t_b/\omega_B = 1$. (b) Dressed single atom (Jaynes-Cummings model).

## V. SOME OTHER POTENTIAL EXPERIMENTAL IMPLEMENTATIONS

### A. Semiconductor 2D-Heterostructures

The next promising candidate for the experimental implementation of the model considered above is the low-dimensional semiconducting heterostructure (see Fig. 2) under femtosecond intersubband excitation [50]. For example, the sample consisting of 51 GaAs quantum wells separated by barriers was implemented in [50]. A coherent excitation of the sample was created by a femtosecond pulse with a center frequency resonant to the intersubband transition. The resonant line at 30THz was homogeneously broadened with coherence time 320Fs. The amplitude of incident field was varied inside of area (5-50) kV/cm. As a result, there were observed coherent subpicosecond RO and manipulated in a wide range by varying the strength of the coherent driving field. The measurements [50] were qualitatively agreed with the predictions of the simplest model based on Maxwell-Bloch equations for non-interactive two-level systems. Such conventional simplifications as RWA and omitting of all types of damping have been used. The simulation of the pulse driving field was having done by the Rabi-frequency varying $\Omega_R = \Omega_R(t)$.

Here, we add the dc field for manipulation by atomic chain via BO. For application of our model to the potential experiments with 2D-heterostructures it is necessary to describe the driving quantum light as a short transient normally incident to the heterostructure. The driving process in this case is described by the instantaneous coupling coefficient $g$. The field quantization should be modified following Appendix C. The quantum properties of light in the wavepacket are dictated by the special pair of bosonic creation-annihilation operators, which allows to rewrite the dynamic equations (7), (8) as

$$i\hbar \frac{\partial a_{p,n}}{\partial t} = \left(\delta\varepsilon - \mathbf{E}_{dc}\cdot e\mathbf{R}_p\right)a_{p,n} + t_a a_{p+1,n} + t_a^* a_{p-1,n} - g(t)\sqrt{n+1}\,b_{p,n+1}, \qquad (36)$$

$$i\hbar \frac{\partial b_{p,n+1}}{\partial t} = -\left(\delta\varepsilon + \mathbf{E}_{dc}\cdot e\mathbf{R}_p\right)b_{p,n+1} + t_b b_{p+1,n+1} + t_b^* b_{p-1,n+1} - g(t)\sqrt{n+1}\,a_{p,n}, \qquad (37)$$

where $g(t) = \mathbf{d}_{ab} \cdot \mathbf{u}_0(t)/\hbar$, $\mathbf{u}_0(t)$ is a slow envelope of the driving pulse.

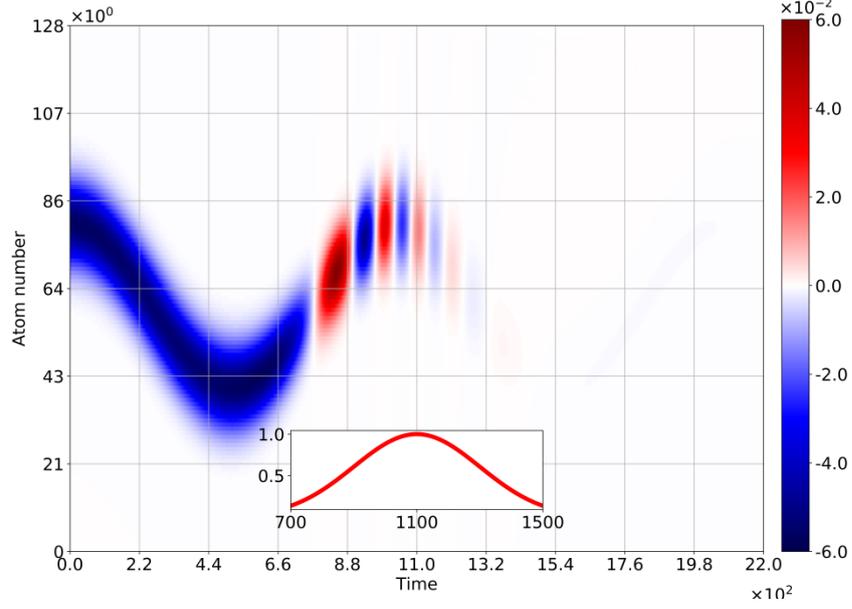

FIG. 16. Space-time distribution of inversion density for pulse light in the coherent state with $<n>=25$. All frequencies are normalized to the frequency of quantum transition $\omega_0$. The number of atoms $N=128$, Rabi-frequency $\Omega_{\langle n \rangle} = 0.2$, electron initial Gaussian with $a_0 = 1, b_0 = 0$, $u_a = 80, \sigma_a = 10, k_a = 0$, $t_a/\omega_B = t_b/\omega_B = 10$. Insert: the envelope of the light pulse in the time scale matched with the main picture.

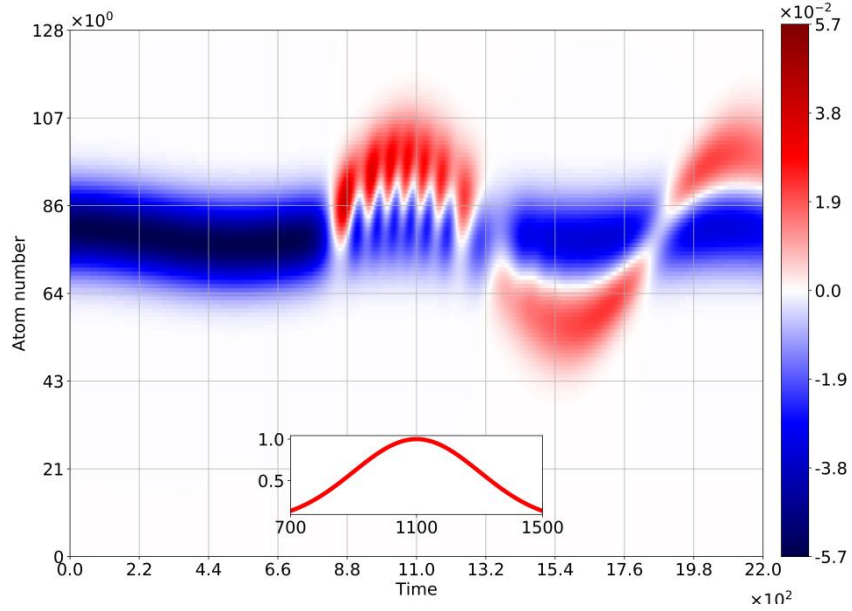

FIG. 17. Space-time distribution of inversion density for pulse light in the coherent state with $<n>=25$. All frequencies are normalized to the frequency of quantum transition $\omega_0$. The number of atoms $N=128$, Rabi-frequency $\Omega_{\langle n \rangle} = 0.2$, electron initial Gaussian with $a_0 = 1, b_0 = 0$, $\omega_B = 0.02$, $u_a = 80, \sigma_a = 10, k_a = 0$, $t_a/\omega_B = 10, t_b/\omega_B = 1$. Insert: the envelope of the light pulse in the time scale matched with the main picture.

The driving with electromagnetic pulse leads to the chain dynamics qualitatively different from the case of monochromatic light. Figures 16,17 depict the temporal-spatial evolution of inversion for different relations between barrier penetrations at 1S and 2S levels. Figure 16 shows the case of identical penetrations. One finds BO over the pulse duration. The oscillations are modulated with Rabi frequency and stop with the pulse disappearing at zero inversion. The behavior becomes quite different for the case of 2S penetration strongly exceed 1S one. (Fig. 17). No oscillations exist before the pulse appearance ( $t < 0.9$ ). The reason for this is a weak value of penetration at 1S level. The switching of pulse creates both RO and BO. The RO existence is limited by the pulse duration ( $t < 1.2$ ), while BO continues even after the pulse switching off. It is a result of tunneling at the 2S-level on which the penetration is much higher. The charge existence at the 2S-level is a result of resonant pumping by the optical pulse. Thus, we can consider the resonant pumping as a tool of BO triggering and speak about photon-assisted BO.

## B. Josephson junction

The next promising candidate for experiment implementation is a single Josephson junction embedded in an inductive environment. As it was shown in Refs.53-56, the voltage-biased Josephson junction exhibits the temporal dynamics equal to the motion of a fictitious charged particle with inertia provided by the inductance. This particle moves in a potential which also has a periodic part and a linear tilt. On the other hand, a Josephson junction strongly coupled to the microwave resonator (Fig. 3) in the resonant regime may be considered as an artificial two-level atom ("fluxonium" [54]). It is described by the Jaynes-Cummings model [54] and exhibits RO dynamics [54]. Here, we will focus on the case of Josephson junction simultaneously driven by dc voltage and quantum light in the regime of small detuning. We will show that the junction parameters correspondent the applicability area of the model investigated before are really reachable. Therefore, we will start our analysis from identifying the parameters of Josephson junction in terms of this model.

The temporal dynamics of entire circuit may be described using so-called quasicharge representation [53,55]. Let us first consider the single junction, which standard form reads $\hat{H}_0 = \left(\hat{Q}^2/2C_J\right) - E_J \cos\hat{\varphi}_J$, where $\hat{Q} = (2e/i)\partial/\partial\hat{\varphi}_J$, $\hat{\varphi}_J$ are operators of junction charge and phase difference across the junction, respectively. It describes the particle of mass $C_J$ moving in the periodic potential [55] ($C_J$ is a junction capacitance). The eigenfunctions of this Hamiltonian are Floquet-Bloch modes with condition of quasi-periodicity $\Psi_{\alpha,Q}(\varphi) = P_{\alpha,Q}(\varphi)e^{i\varphi Q/2e}$, where $P_{\alpha,Q}(\varphi) = P_{\alpha,Q}(\varphi + 2\pi)$ is periodic function, $\alpha$ is a number of the gap. These functions describe the motion of the quasiparticles, which is similar to the motion of electrons in crystal. The phase $\varphi$ and quasicharge $Q$ play role of conjugated spatial coordinate and quasi-momentum for electrons in crystals, respectively. The energies of the two first modes $\varepsilon_{0,1}(Q)$ are separated by the gap, which in the regime of interest here ( $E_J \gg E_{C_J}, E_{C_J} = e^2/2C_J$ ) is of the order of plasma frequency $\omega_p = \sqrt{8E_J E_{C_J}}/\hbar$. Taking $E_J/h = 30$ GHz, and $E_{C_J}/h = 3$, we have $\omega_p = 150$ GHz. Next, in the limit $E_J \gg E_{C_J}$ the energy bands $\varepsilon_{0,1}(Q)$ are purely sinusoidal [53] with the bandwidths of the order of [55]

$$\Delta_0 = 16\sqrt{E_{C_J}E_J/\pi}\left(E_J/2E_{C_J}\right)^{1/4} e^{-\sqrt{8E_J/E_{C_J}}} \tag{38}$$

(the typical values are $\Delta_0 \approx E_C/2 = 1,5$ GHz). The situation is the same in the case of atomic chain considered above (see equation (13) in the case $g_n = 0$). Thus, we are able to establish the correspondence $2t_{a,b} \leftrightarrow \Delta_0$.

The additional series inductance and switching on the bias voltage transforms Hamiltonian $\hat{H}_0$ to

$$\hat{H}_A = \hat{Q}^2/2C_J + E_{L_J}(\phi_b - \hat{\varphi}_J)^2/2 + E_J \cos(\hat{\varphi}_J), \tag{39}$$

where $E_{L_J} = (\hbar/2e)^2 L_2^{-1}$ is the constant of inductance energy, $\phi_b$ is the total phase difference across the circuit, which is determined by the bias voltage $V_b$, such that $\dot{\phi}_b = 2eV_b/\hbar$ [53].

The case under consideration has a simple physical interpretation based on the charge-phase duality [53]. The voltage-biased junction with inductive environment and the current-biased junction shunted by a capacitor $C_J$ are dual to each other when exchanging the role of quasicharge and phase. The dynamics of the circuit shown in Fig.3 is equal to the intraband motion of fictitious charged particle in the tilted periodic potential. The tilt leads to BO of the particle in the quasicharge space, which allows identifying the Bloch-frequency as dual plasma frequency $\omega_B \Leftrightarrow \omega_c = \pi(\Delta_0 L_2)^{1/2}/e$.

The inductive strong coupling of a Josephson junction with microwave resonator leads to its dressing with microwave photons. It corresponds to the periodic emission-absorption of photons and interband transitions by the fictitious charged particle (RO). Thus, the qualitative picture of junction behavior is similar to the BO of dressed electrons considered above: the periodic interband transitions (RO of fluxonium) are accompanied by the charge oscillations through the bias-voltage driving. For formal identification of the coupling parameters of the system "junction + microwave resonator" it is necessary to analyze in more detail the total Hamiltonian. It reads

$$\hat{H} = \hat{H}_A + \hat{H}_{ph} + \hat{H}_{I,ph} + \hat{H}_{I,dc} \tag{40}$$

where $\hat{H}_A$ is a Hamiltonian of the Josephson junction in an inductive environment, $\hat{H}_{ph}$ is Hamiltonian of microwave photons in the resonator, $\hat{H}_{I,ph}$ is a Hamiltonian of junction-cavity interaction, $\hat{H}_{I,dc}$ is a support of biased voltage. For further analysis it is convenient to use the secondary quantization technique. The Hamiltonian of photons reads

$$\hat{H}_{ph} = \hat{\Phi}_r^2/2L_r + \hat{Q}_r^2/2C_r \tag{41}$$

where $\hat{Q}_r = i\sqrt{\frac{\hbar}{2Z_r}}(\hat{a} - \hat{a}^+)$, $\hat{\Phi}_r = i\sqrt{\frac{\hbar Z_r}{2}}(\hat{a} + \hat{a}^+)$, $\hat{a}, \hat{a}^+$ are creation-annihilation operators for microwave photons, which satisfy the conventional bosonic commutative relations. By following the concept of quasicharge-phase duality, the Fourier-Bloch states $\Psi_{\alpha,Q}(\varphi)$ exchanged by the Wannier functions (see Appendix A) and Hamiltonian is written in the Wannier-basis. We denote the two first states with quasicharge $Q$ as $|e_Q\rangle$ and $|g_Q\rangle$ and introduce the raising operator $\hat{\sigma}_Q = |e_Q\rangle\langle g_Q|$ and $\hat{\sigma}_Q^+ = |g_Q\rangle\langle e_Q|$. The atomic Hamiltonian may be rewritten as $\hat{H}_A = \hbar\omega_p \sum_Q \hat{\sigma}_{Qz}/2$, where $\hat{\sigma}_{Qz}$ is Pauli inversion matrix.

For junction-cavity interaction we have [54]

$$\hat{H}_I = \frac{2e}{\hbar} G\hat{\Phi}_r \phi_{eg} \sum_Q (\hat{\sigma}_Q^+ + \hat{\sigma}_Q^-) \tag{42}$$

where $G \approx (\hbar/2e)^2 L_1 (L_2 L_r)^{-1}$, $\phi_{eg} = \langle g_q | \hat{\sigma}^+ | e_q \rangle$ is the matrix element of effective dipole moment. The Hamiltonian is written in rotating-wave approximation. It is equal to the Hamiltonian (3) if the coupling coefficient identified with the value

$$g \Leftrightarrow \frac{|\phi_{eg}|^2}{\sqrt{2}} \sqrt{\frac{R_0}{2\pi Z_r} \frac{L_1}{L_2}} \omega_p \qquad (43)$$

where $R_0$ is the quantum of resistance. The validity of rotating-wave approximation implies all relevant energies to be smaller than the plasmonic frequency of the Josephson junction and resonant frequency of resonator. In particular, we impose the characteristic energy $E_{L_J}$ associated with the inductance $L_2$ to be smaller than $\hbar\omega_p$. It corresponds to the values $L_2 \approx 200$ nH, which are quite reachable [56]. Second, we need also $g \ll \omega_p$. For standard value $Z_r = 50$ k$\Omega$ [56] we have $L_1/L_2 \approx 10^{-3}$ for this condition been imposed. The coherence time and energy relaxation time both are of the order $\cong \mu$s [56]. Thus, the number of BO flops is high enough for their experimental observation in the fluxonium atom in microwave frequency range. The real cases may be described by lossless model considered above.

## VI. CONCLUSION AND OUTLOOK

In this paper, we developed the model of the chain of two-level artificial atoms manipulated simultaneously by the dc field and single-mode quantum light in the strong coupling regime. Atom-light interaction assumed to be resonant. The cases of monochromatic light and light pulse have been considered. Both, electronic and photonic characteristics of the oscillation process, have been studied, such as: i) inversion density; ii) tunneling current density; iii) distribution of photon probabilities; iv) mean number of photons; v) photon number variance; vi) Neumann entropy of light. The different types of the initial states of the light have been considered: i) coherent state; ii) vacuum photonic state; iii) Fock-state entangled with the electron wave packet. The Gaussian wavepackets were chosen as initial states of atoms.

The model is universal with respect to the physical origin of artificial atoms and frequency ranges of atom-light interaction. The model was adapted to the semiconductor 2D-heterostructures (THz frequencies), semiconductor quantum dots (optical range), and Josephson junctions (microwaves). The initial data for numerical simulations are taken from recently published experiments.

The dynamical equations have been studied both analytically and numerically. The idea for analytical simplification is based on the separate description of electrons dressing with photons and quasiclassical motion of their geometrical center driven by dc field. The analytical solutions are in good qualitative agreement with numerical simulations. Our model is based on such conventional simplifications as neglect of damping and RWA. Their validity for considered systems was supported by numerical estimations taken from experimental data.

The following conclusions are emerged from our studies:

1). The case of initial coherent state of the light exhibits the collapse-revival picture, which drifts over the chain (collapses and revivals placed in the different spatial areas). In contrast with Jaynes-Cummings model, the collapse-revival picture is modulated with Bloch frequency;

2). In the case of initial vacuum state of light, the photon emission and absorption occurs with Bloch frequency, instead of Rabi-frequency in Jaynes-Cummings model. The photonic probabilities are mainly modulated with Bloch frequency, while the contribution of Rabi-components is rather slight. BO strongly squeezes the vacuum state of light entangled with electronic wavepackets;

3). The electron-photon entanglement dramatically modifies the tunnel current behavior. It becomes modulated agreed with the collapse-revival picture for the case of coherent state of light, and periodically modulated by RO for the case of initial photonic vacuum;

To conclude, the main result of the paper is the novel effect of influence of BO and quantum statistical properties of light on each other. It is counterintuitive because of the strongly different frequency ranges for such types of oscillations existence. The reason for it is an entanglement of electronic and photonic states in the system been considered.

The obtained results allow to control of quantum-statistical properties of light via adiabatically turning dc field. They are promising for applications in quantum optics, quantum informatics, quantum computing. The mutual modulation of low-frequency BO and high-frequency inter-band transitions produces the new types of spectral lines in tunneling current and optical polarization. It opens the promising ways in spectroscopy of nanodevices in THz and optical frequency ranges. These problems may be considered as directions for future research activity.

## ACKNOWLEDGMENTS

G.Ya. acknowledges support from the project FP7-PEOPLE-2013-IRSES-612285 CANTOR.

## APPENDIX A: DERIVATION OF HAMILTONIAN IN THE WANNIER-FOCK BASIS

The Hamiltonian of conductive electron movement in the chain associated with the interatomic tunneling, which is described by the periodic potential $V(\mathbf{r})$, is

$$\hat{H}_0 = \frac{\hat{\mathbf{p}}^2}{2m} + V(\mathbf{r}) \tag{A1}$$

The eigenmodes of Hamiltonian (A1) are two Bloch modes correspondent to valence (*b*) and conductive (*a*) zones and denoted as $\psi_{\alpha,h}(\mathbf{r}) = |\alpha, h\rangle$, $\alpha = a, b$, $h$ being a scalar quasi-momentum directed along the chain. Because the electrons are strongly confined inside the atoms, we will use as a basis Wannier functions $\phi_{\alpha,\mathbf{R}_j}(\mathbf{r}) = |\alpha_j\rangle$ [69], defined as a linear combinations of Bloch modes

$$\phi_{\alpha,\mathbf{R}_j}(\mathbf{r}) = \frac{1}{\sqrt{N}} \sum_h e^{-ih(\mathbf{e}\cdot\mathbf{R}_j)} \psi_{\alpha,h}(\mathbf{r}) \tag{A2}$$

Let us mention some properties of Wannier functions, which are important for our analysis. Wannier function $\phi_{\alpha,\mathbf{R}_j}(\mathbf{r})$ is strongly localized inside the *j*-th atom. Inversion of Eq. (A2) gives the expression of Bloch states in the terms of Wannier functions:

$$\psi_{\alpha,h}(\mathbf{r}) = \frac{1}{\sqrt{N}} \sum_{\mathbf{R}_j} e^{ih(\mathbf{e}\cdot\mathbf{R}_j)} \phi_{\alpha,\mathbf{R}_j}(\mathbf{r} - \mathbf{R}_j) \tag{A3}$$

Wannier functions are satisfying the orthogonality relation $\langle \alpha_j | \alpha_l \rangle = \delta_{\alpha\alpha'}\delta_{jl}$.

Projecting $\hat{H}_0$ on the Bloch modes and using (A3), we obtain the dispersion law for free-tunneling electron in the next form:

$$\varepsilon_\alpha(h) = \langle \alpha, h | \hat{H}_0 | \alpha, h \rangle = \frac{1}{N} \sum_{j,l} e^{-ih\mathbf{e}\cdot(\mathbf{R}_j - \mathbf{R}_l)} \langle \alpha_j | \hat{H}_0 | \alpha_l \rangle \tag{A4}$$

Denoting

$$\langle \alpha_j | \hat{H}_0 | \alpha_l \rangle = \begin{cases} \varepsilon_\alpha(0), & j = l \\ t_\alpha(\mathbf{R}_j - \mathbf{R}_l), & j \neq l \end{cases} \tag{A5}$$

we can rewrite (A4) in the following form:

$$\varepsilon_\alpha(h) = \varepsilon_\alpha(0) + \frac{1}{N} \sum_{\substack{j,l \\ j \neq l}} e^{-ih\mathbf{e}\cdot(\mathbf{R}_j-\mathbf{R}_l)} t_\alpha(\mathbf{R}_j - \mathbf{R}_l) \tag{A6}$$

The Hamiltonian (A1) may be presented with the Wannier basis in the form of a block diagonal matrix

$$\hat{H}_0 = \begin{pmatrix} \hat{H}_{0a} & 0 \\ 0 & \hat{H}_{0b} \end{pmatrix} \tag{A7}$$

with diagonal-on elements

$$\hat{H}_{0\alpha} = \varepsilon_\alpha(0) \sum_j |\alpha_j\rangle\langle\alpha_j| + \sum_{\substack{j,l \\ j \neq l}} t_\alpha(\mathbf{R}_j - \mathbf{R}_l)|\alpha_l\rangle\langle\alpha_j| \tag{A8}$$

where $\alpha = a,b$. The tunneling coupling exist only between the neighboring atoms (tight-binding approximation). Thus, we omit for brevity the spatial arguments in tunneling matrix elements: $t_{a,b}(\mathbf{R}_j - \mathbf{R}_{j+1}) = t_{a,b}$, $t_{a,b}(\mathbf{R}_j - \mathbf{R}_{j-1}) = t_{a,b}^*$. The Hamiltonian (A7) subject to (A8) reads

$$\hat{H}_0 = \varepsilon_b(0) \sum_j |b_j\rangle\langle b_j| + \varepsilon_a(0) \sum_j |a_j\rangle\langle a_j| +$$
$$\sum_j (t_b|b_j\rangle\langle b_{j+1}| + t_b^*|b_j\rangle\langle b_{j-1}|) + \tag{A9}$$
$$\sum_j (t_a|a_j\rangle\langle a_{j+1}| + t_a^*|a_j\rangle\langle a_{j-1}|)$$

Hamiltonian (A9) in the secondary quantization form is equivalent to the sum of intra-level term and tunneling Hamiltonian.

The total Hamiltonian may be presented as $\hat{H} = \hat{H}_0 + \hat{H}_{ph} + \hat{H}_{I,ph} + \hat{H}_{I,dc}$, where the second component $\hat{H}_{ph} = \hbar\omega \hat{a}^+ \hat{a}$ corresponds to the free single-mode photons motion, and $\hat{a}, \hat{a}^+$ are creation and annihilation operators of photons. We describe the atom-photon interaction in the dipole approximation by the third term $\hat{H}_{I,ph} = -e\hat{\mathbf{E}}(\mathbf{r},t)\cdot\hat{\mathbf{r}}$ [26], where

$$\hat{\mathbf{E}} = \mathbf{e}\sqrt{\frac{\hbar\omega}{2\varepsilon_0 V}}(\hat{a} + \hat{a}^+) \tag{A10}$$

is the single-mode quantum light operator of electric field, *V* is the normalizing volume, *e* is the unit polarization vector.

The quantum state of light will be described in terms of photons number states (Fock-states) $|m\rangle$. The basis states should be used in our analysis correspond to the independent photons number states and Wannier electrons $|\alpha_j, m\rangle$ (Wannier-Fock basis). Let us proceed to the calculation of matrix elements

$$H_{I,ph}^{mn,lj,\alpha\alpha'} = \langle m,\alpha_l|\hat{H}_{I,ph}|\alpha_j',n\rangle = -e\sqrt{\frac{\hbar\omega}{2\varepsilon_0 V}}\langle\alpha_l|\mathbf{e}\cdot\hat{\mathbf{r}}_j|\alpha_j'\rangle\langle m|(\hat{a}+\hat{a}^+)|n\rangle \tag{A11}$$

Let us shift the origin for a given $j$ to the center of the $j$-th atom via the relation $\hat{\mathbf{r}}_j = \mathbf{R}_j \hat{I} + \hat{\mathbf{r}}'$. Using the Wannier functions orthogonality for different electron bands, we have $\langle \alpha_j | \hat{\mathbf{r}}_j | \alpha'_j \rangle = \mathbf{R}_j \delta_{\alpha\alpha'} + \langle \alpha_j | \hat{\mathbf{r}}' | \alpha'_j \rangle$. Together with orthogonality of the Fock states it makes (A11)

$$H_{I,ph}^{mn,jl,\alpha\alpha'} = -e\sqrt{\frac{\hbar\omega}{2\varepsilon_0 V}} \left( \sqrt{n}\delta_{m,n-1} + \sqrt{n+1}\delta_{m,n+1} \right) \begin{cases} \left[ (\mathbf{e}\cdot\mathbf{R}_j)\delta_{\alpha\alpha'} + (\mathbf{e}\cdot\mathbf{d}_{\alpha\alpha'}) \right], j=l \\ \delta_{\alpha\alpha'} \tilde{d}_{j,\alpha}, l=j+1 \\ 0, others \end{cases} \quad (A12)$$

Here $\mathbf{d}_{\alpha\alpha'} = \langle \alpha_j | \mathbf{r}_j | \alpha'_j \rangle$ is the matrix of the dipole moments, $\tilde{d}_{j,\alpha} = \langle \alpha_j | (\mathbf{e}\cdot\mathbf{r}_j) | \alpha_{j+1} \rangle$. The second term at the top line characterizes quantum transitions in the separate atom. For real atoms (excluding hydrogen) only off-diagonal elements $\mathbf{d}_{\alpha\alpha'}$ are non-zero due to the inversion symmetry [70]. For some types of artificial atoms (such as quantum dots) the inversion symmetry is not obligatory. In this case on-diagonal elements of dipole matrix may be non-zero [37], which corresponds to the special type of intra-level motion [37]. The middle line in (A12) corresponds to the interaction of neighboring atoms stimulated by photons (photon-assisted tunneling over the same level).

The last term $\hat{H}_{I,dc} = -e\mathbf{E}_{dc}(\mathbf{r},t) \cdot \hat{\mathbf{r}}$ describes the dipole interaction of the chain with the dc field. Its matrix elements are

$$H_{I,dc}^{mn,lj,\alpha\alpha'} = \langle m, \alpha_l | \hat{H}_{I,dc} | \alpha'_j, n \rangle = -e\delta_{mn}\mathbf{E}_{dc} \cdot \langle \alpha_l | \hat{\mathbf{r}} | \alpha'_j \rangle \quad (A13)$$

Using the orthogonality of Wannier states, we obtain

$$H_{I,dc}^{mn,jl,\alpha\alpha'} = -e\delta_{mn}\mathbf{E}_{dc} \begin{cases} \mathbf{R}_j \delta_{\alpha\alpha'} + \mathbf{d}_{\alpha\alpha'}, j=l \\ \delta_{\alpha\alpha'} \tilde{d}_{j,\alpha} \mathbf{e}, l=j+1 \\ 0, others \end{cases} \quad (A.14)$$

The first term in the top line corresponds to BO, the second one describes Stark-effect [70].

### APPENDIX B: DERIVATION OF FORMULA FOR TUNNELING CURRENT

For calculating the tunneling current we introduce the operator of the particle number in the $j$-th atom $\hat{N}_j = |a_j\rangle\langle a_j| + |b_j\rangle\langle b_j|$ and formulate the equation of continuity

$$\hat{J}_j - \hat{J}_{j-1} = -e\frac{d\hat{N}_j}{dt} \quad (B1)$$

where $\hat{J}_j$ is the current density operator in the $j$-th atom and the left-hand part in (A1) is a discrete analog of divergence in 1D-case. These operators act only to the electronic states; therefore all relations of this Appendix are given in terms of Wannier states. Using Heisenberg equation for the operator $\hat{N}_j$, we rewrite (B1) in the form

$$\hat{J}_j - \hat{J}_{j-1} = -i\frac{e}{\hbar}\left[\hat{H}, \hat{N}_j\right] \quad (B2)$$

The tunneling currents over excited and background energy levels are independent, thus $\hat{J}_{Tunneling,j} = \hat{J}_{Tunneling,j}^{(a)} + \hat{J}_{Tunneling,j}^{(b)}$. Using (B2) we obtain

$$\hat{J}^{(a,b)}_{\text{Tunneling},j} - \hat{J}^{(a,b)}_{\text{Tunneling},j-1} = -i\frac{e}{\hbar}\left[\hat{H}^{(a,b)}_{\text{Tunneling}}, \hat{N}_j\right] \quad (B3)$$

where

$$\hat{H}^{(a)}_{\text{Tunneling}} = \sum_j (t_a |a_j\rangle\langle a_{j+1}| + t_a^* |a_j\rangle\langle a_{j-1}|) \quad (B4)$$

is a component of Hamiltonian (2), chargeable for the tunneling at the excited level (the similar equation may be written for the Hamiltonian $\hat{H}^{(b)}_{\text{Tunneling}}$). Using (B4), we calculate the commutator in the right-hand part of (B3) and obtain

$$\hat{J}^{(a)}_{\text{Tunneling},j} - \hat{J}^{(a)}_{\text{Tunneling},j-1} = -iet_a \left(|a_{j-1}\rangle + |a_{j+1}\rangle\right)\langle a_j| + H.c. \quad (B5)$$

It corresponds to the operator of the tunneling current at the excited level

$$\hat{J}^{(a)}_{\text{Tunneling},j} = -iet_a |a_j\rangle\langle a_{j+1}| + H.c. \quad (B6)$$

The observable value of the tunneling current is

$$J^{(a)}_{\text{Tunneling},j}(t) = \left\langle \hat{J}^{(a)}_{\text{Tunneling},j} \right\rangle = -iet_a a_j^*(t) a_{j-1}(t) + c.c. \quad (B7)$$

Using the approximation $a_{j-1}(t) - a_j(t) \approx \left(a_{j+1}(t) - a_{j-1}(t)\right)/2$, and adding the similar support of the ground level, we obtain the relation (10). This relation corresponds to the well-known definition of the probability flow $\mathbf{j} = i\hbar\left(\Psi \text{grad}\Psi^* - c.c.\right)/2m$ in 3D continuous case [70].

**APPENDIX C: QUANTIZATION OF ELECTROMAGNETIC FIELDS IN WAVEPACKETS**

For the secondary quantization in the wavepacket we use the method proposed by O. Keller [71]. First we consider the linearly polarized electric field to have the spatial-temporal dependence for a 1D cavity resonator [26]. We assume the cavity formed by two parallel perfectly conductive planes distanced on the length $L$. The positive-frequency part of electric field operator is given [26]

$$\hat{\mathbf{E}}^{(+)}(x,t) = \mathbf{e}\sum_\nu \sqrt{\frac{\hbar\omega_\nu}{2\varepsilon_0 V}} \hat{a}_\nu e^{i(k_\nu x - \omega_\nu t)} \quad (C1)$$

where $V$ is normalization volume, $k_\nu = \nu\pi/L$, $\nu$ is an integer value, $\omega_\nu = k_\nu c$.

Next, following [71] we introduce the special complete system of basis classical wavepackets $\mathbf{u}_\alpha(x,t)$, everyone of which is given by the superposition of cavity modes

$$\mathbf{u}_\alpha(x,t) = \mathbf{e}\sum_\beta \sqrt{\frac{\hbar\omega_\beta}{2\varepsilon_0 V}} B_{\alpha\beta} e^{i(k_\beta x - \omega_\beta t)} \quad (C2)$$

where $B_{\alpha\beta}$ are elements of unit matrix with conventional unitary condition $\sum_\gamma B_{\gamma\alpha}B_{\gamma\beta}^* = \delta_{\alpha\beta}$, $\delta_{\alpha\beta}$ is the Kronecker delta. Equation (C1) using (C2) may be rewritten as

$$\hat{\mathbf{E}}^{(+)}(x,t) = \mathbf{e}\sum_{v}\sum_{v'}\delta_{vv'}\sqrt{\frac{\hbar\omega_v}{2\varepsilon_0 V}}\hat{a}_{v'}e^{i(k_v x - \omega_v t)} =$$

$$\mathbf{e}\sum_{v}\sum_{v'}\sum_{\gamma}B_{\gamma v'}B^*_{\gamma v}\sqrt{\frac{\hbar\omega_v}{2\varepsilon_0 V}}\hat{a}_{v'}e^{i(k_v x - \omega_v t)} = \quad \text{(C3)}$$

$$\sum_{v}\sum_{\gamma}B_{\gamma v'}\hat{a}_{v'}\mathbf{u}_\gamma(x,t)$$

Using (C2) and unitary condition, we obtain

$$\hat{\mathbf{E}}^{(+)}(x,t) = \sum_{\gamma}\hat{c}_\gamma \mathbf{u}_\gamma(x,t) \quad \text{(C4)}$$

where $\hat{c}_\gamma = \sum_{v} B_{\gamma v}\hat{a}_v$ is a new set of creation-annihilation operators, satisfying Bose commutative relations $\left[\hat{c}_\alpha, \hat{c}^+_\beta\right] = \delta_{\alpha\beta}$. The basis wavepackets are non-orthogonal:

$$\int_V \mathbf{u}_p(x,t)\cdot\mathbf{u}^*_q(x,t)dV = \sum_{\alpha}\frac{\hbar\omega_\alpha}{2\varepsilon_0}B_{p\alpha}B^*_{q\alpha} \quad \text{(C5)}$$

The field Hamiltonian is

$$\hat{H}_{ph} = \varepsilon_0\int_V \left\{\hat{\mathbf{E}}^{(+)}(x,t)\cdot\hat{\mathbf{E}}^{(-)}(x,t) + \hat{\mathbf{E}}^{(-)}(x,t)\cdot\hat{\mathbf{E}}^{(+)}(x,t)\right\}dV =$$

$$\varepsilon_0\sum_{p}\sum_{q}\left(\hat{c}_p\hat{c}^+_q + \hat{c}^+_q\hat{c}_p\right)\int_V \mathbf{u}_p(x,t)\cdot\mathbf{u}^*_q(x,t)dV = \quad \text{(C6)}$$

$$\sum_{\alpha}\frac{\hbar\omega_\alpha}{2}\sum_{p}\sum_{q}B_{p\alpha}B^*_{q\alpha}\left(\hat{c}_p\hat{c}^+_q + \hat{c}^+_q\hat{c}_p\right)$$

Due to the property (C5), the energies of different modes are mixed via their mutual interference.

For applying the single mode approximation, some additional simplifications should be done. Let us assume that we consider the non-monochromatic field with the narrow frequency spectrum, which localized in the vicinity of central frequency $\omega_\alpha \approx \tilde{\omega}$. In this case the different modes $\mathbf{u}_\alpha(x,t)$ become approximately orthogonal and their interference disappears. The model of the wavepacket in the first order of dispersion theory gives

$$\mathbf{u}_p(x,t) = \mathbf{u}(x,t) \approx \mathbf{u}_0\left(t - \frac{x}{v_{gr}}\right)e^{i(k(\tilde{\omega})x - \tilde{\omega}t)} \quad \text{(C7)}$$

where $\mathbf{u}_0(t)$ is a slow envelope, the exponential factor gives the high-frequency filling, $v_{gr}$ is a group velocity. The approximate field Hamiltonian and electric field operator read

$$\hat{H}_{ph} \approx \hbar\tilde{\omega}\left(\hat{c}^+\hat{c} + \frac{1}{2}\right) \quad \text{(C8)}$$

$$\hat{\mathbf{E}}(x,t) \approx \mathbf{u}_0\left(t - \frac{x}{v_{gr}}\right)\hat{c}e^{i(k(\tilde{\omega})x - \tilde{\omega}t)} + \text{H.c.} \quad \text{(C9)}$$

respectively, where $\hat{c}^+, \hat{c}$ are annihilation-creation operators for photon in the wavepacket. The coupling factor $g = \mathbf{d}_{ab} \cdot \mathbf{u}_0(t)/\hbar$ becomes time-dependent, what allows us to consider the case of driving light in the form of the rather long laser pulse (assumed that the pulse duration strongly exceeds the period of high-frequency filling, which guarantees the RWA validity in our model).